\documentclass[aps,superscriptaddress,pra,twocolumn]{revtex4-1}

\usepackage{amssymb,amsmath,amsfonts,mathrsfs,bm, bbm, graphicx, epsfig, epstopdf}
\usepackage{braket}
\usepackage{hyperref}
\usepackage{color}
\newcommand\mycom[2]{\genfrac{}{}{0pt}{}{#1}{#2}}

\begin{document}

\title{Analyzing photon-count heralded entanglement generation between solid-state spin qubits by decomposing the master equation dynamics}

\author{Stephen C. Wein}
\email{wein.stephen@gmail.com}
\affiliation{Institute for Quantum Science and Technology, and Department of Physics \& Astronomy, University of Calgary, 2500 University Drive NW, Calgary, Alberta T2N 1N4, Canada}

\author{Jia-Wei Ji}
\affiliation{Institute for Quantum Science and Technology, and Department of Physics \& Astronomy, University of Calgary, 2500 University Drive NW, Calgary, Alberta T2N 1N4, Canada}

\author{Yu-Feng Wu}
\affiliation{Institute for Quantum Science and Technology, and Department of Physics \& Astronomy, University of Calgary, 2500 University Drive NW, Calgary, Alberta T2N 1N4, Canada}

\author{Faezeh Kimiaee Asadi}
\affiliation{Institute for Quantum Science and Technology, and Department of Physics \& Astronomy, University of Calgary, 2500 University Drive NW, Calgary, Alberta T2N 1N4, Canada}

\author{Roohollah Ghobadi}
\affiliation{Institute for Quantum Science and Technology, and Department of Physics \& Astronomy, University of Calgary, 2500 University Drive NW, Calgary, Alberta T2N 1N4, Canada}

\author{Christoph Simon}
\email{christoph.simon@gmail.com}
\affiliation{Institute for Quantum Science and Technology, and Department of Physics \& Astronomy, University of Calgary, 2500 University Drive NW, Calgary, Alberta T2N 1N4, Canada}

\begin{abstract}
    We analyze and compare three different schemes that can be used to generate entanglement between spin qubits in optically-active single solid-state quantum systems. Each scheme is based on first generating entanglement between the spin degree of freedom and either the photon number, the time bin, or the polarization degree of freedom of photons emitted by the systems. We compute the time evolution of the entanglement generation process by decomposing the dynamics of a Markovian master equation into a set of propagation superoperators conditioned on the cumulative detector photon count. We then use the conditional density operator solutions to compute the efficiency and fidelity of the final spin-spin entangled state while accounting for spin decoherence, optical pure dephasing, spectral diffusion, photon loss, phase errors, detector dark counts, and detector photon number resolution limitations. We find that the limit to fidelity for each scheme is restricted by the mean wavepacket overlap of photons from each source, but that these bounds are different for each scheme. We also compare the performance of each scheme as a function of the distance between spin qubits.
\end{abstract}

\maketitle

\section{Introduction}
\label{sec:introduction}

Photon-mediated entanglement generation between quantum systems is important for implementing quantum repeaters \cite{briegel1998quantum, duan2001long,sangouard2009quantum,sangouard2011quantum,asadi2018quantum,rozpkedek2019near} and distributed quantum computing protocols \cite{lim2005repeat, benjamin2009prospects}, which may one day be used to build a quantum internet \cite{kimble2008internet,simon2017internet,wehner2018internet}. To achieve this for systems that emit visible or near-infrared photons, it is convenient to use pulsed schemes that herald entanglement by the detection of single photons \cite{cabrillo1999creation,barrett2005efficient,saucke2002thesis,simon2003robust,feng2003entangling,feng2003entangling}. Such schemes have already been implemented using atomic ensembles \cite{chou2005measurement,chou2007functional}, single trapped atoms \cite{hofmann2012heralded} or ions \cite{moehring2007entanglement,slodivcka2013atom}, quantum dots \cite{delteil2016generation,stockill2017phase}, and defects in diamond \cite{bernien2013heralded,hensen2015experimental}.

Solid-state systems are particularly attractive as a quantum communication platform for their scalability, ease of manufacturing, and potential to integrate with classical information processing hardware \cite{awschalom2018quantum,atature2018material}. However, solid-state systems are subject to dephasing processes that limit the initial amount of generated entanglement between systems as well as their longevity as a quantum memory \cite{benjamin2009prospects}. Spin decoherence can be caused by the interaction of the spin qubit with a surrounding bath of nuclear spins \cite{hanson2007spins} or lattice phonons \cite{golovach2004phonon}. These interactions can randomly flip the spin state of the qubit during or after entanglement generation, or cause a pure-dephasing of the spin coherence. Phonon interactions can also cause homogeneous broadening of the zero-phonon line (ZPL) for solid-state optical transitions \cite{fu2009observation,plakhotnik2015electron}, which degrades the indistinguishability of photons emitted from the quantum system \cite{grange2015cavity}. These decoherence processes limit the amount of spin-photon entanglement and, consequently, the amount of final spin-spin entanglement.

The two most critical figures of merit for entanglement generation are efficiency and fidelity. The efficiency impacts the overall rate of quantum information transfer. For example, the quantum key distribution rate for a repeater protocol is proportional to the entanglement generation efficiency. The fidelity quantifies the quality of entanglement in addition to our knowledge about the state of the system. High-fidelity entanglement is necessary for many quantum information applications. In addition, purification and error correction protocols require minimum fidelity thresholds to be satisfied \cite{dur1999quantum}. 

In this paper, we apply a photon count decomposition to compute the entanglement generation efficiency and fidelity of spin-spin entanglement heralded by photon counting measurements. This approach uses a Liouville-Neumann series \cite{carmichael,horoshko1998multimode, brun2000continuous} to decompose the master equation dynamics into a set of propagation superoperators that describe the spin state evolution conditioned on the cumulative detector photon count during a window of time.

Conditional evolution is the foundation for the quantum trajectories method \cite{carmichael,wiseman1994quantum,daley2014quantum} and is usually applied to reduce the computational complexity of the master equation by solving an effective Schr\"{o}dinger equation or a stochastic equation of the open quantum system \cite{zoller1987quantum,carmichael1989photoelectron,browne2003robust,dodonov2005microscopic,barrett2005efficient,zhang2018quantum,zhang2019heralded}. Recently, the photon count decomposition has been used to compute photon statistics for complicated emitter dynamics \cite{hanschke2018quantum,zhang2018quantum}, to propose a heralded entanglement generation scheme for classically-driven emitters coupled to a waveguide \cite{zhang2019heralded}, and has been connected to the exact emission field state of a quantum emitter coupled to a waveguide \cite{fischer2018particle,fischer2018scattering}, justifying a physical interpretation of quantum trajectories \cite{fischer2018scattering}. It is also intrinsically related to continuous measurement and quantum feedback theories \cite{wiseman1994quantum,wiseman2009quantum}, allowing for the computation of active feedback schemes. As we will show, this decomposition is also a powerful analytic tool to analyze photon count post-selection, or passive feedback \cite{wiseman1994quantum}, schemes under the effects of decoherence. In this context, the photon counting measurement does not affect the average evolution of the open quantum system and the conditional evolution description instead serves to properly describe the collapse of the measured system.

A strength of the photon count decomposition is that it can be used to compute entanglement figures of merit while accounting for a multitude of realistic imperfections for any emitter Markovian master equation, including those with pure dephasing effects. This circumvents modelling the full spin-photon system and subsequently tracing out the photonic modes by instead relying on the input-output relations for the open quantum systems \cite{gardiner1985input}. Furthermore, when considering direct photon counting measurements where there is no local oscillator, the vacuum fluctuations vanish allowing for a direct proportionality between the system operators and the emitted field \cite{carmichael1989photoelectron,wiseman1994quantum}. With this method, we take into account imperfections such as spin flips, spin pure dephasing, optical pure dephasing, spectral diffusion, collection inefficiency, transmission loss, phase errors, detector inefficiency, detector dark counts, and limited detector number-resolving capabilities.

We analyze and compare three different popular spin-spin entanglement generation protocols: (1) via spin-photon number entanglement with a single pulse \cite{cabrillo1999creation}, (2) via spin-time bin entanglement with two sequential $\pi$-pulses \cite{barrett2005efficient}, and (3) via spin-polarization entanglement using an excited $\Lambda$ system \cite{saucke2002thesis,simon2003robust,feng2003entangling}. Each of these require fast resonant pulsed excitation of the quantum systems. For each protocol, we derive the spin-spin conditional states for photon counting measurements to compute expressions for the entanglement figures of merit. We also discuss their relationship to the properties of single-photon emission from the individual quantum systems, such as brightness and mean wavepacket overlap.

We outline our methods in section \ref{sec:methods}. Section \ref{sec:protocols} presents our analytical and numerical results for each protocol. We discuss and compare the protocols in section \ref{sec:discussion} and give our conclusions in section \ref{sec:conclusions}.

\section{Methods}
\label{sec:methods}

\subsection{Photon count decomposition}
\label{subsec:evolution}

Consider a general Markovian master equation
\begin{equation}
\label{meq}
    \dot{\rho} = \mathcal{L}\hat{\rho}(t),
\end{equation}
where $\mathcal{L}$ is the Liouville superoperator that contains all the reversible and irreversible dynamics of the open system. In our analysis, we notate all superoperators using a calligraphic font and we assume that they act on everything situated to their right unless otherwise specified. 

To decompose the master equation dynamics into evolution conditioned on single photon detection, we can rearrange the master equation in the following way:
\begin{equation}
\label{meqC}
    \dot{\rho}=\mathcal{L}_0\hat{\rho}(t) +\sum_{i=1}^N\mathcal{S}_{i}\hat{\rho}(t),
\end{equation}
where $\mathcal{L}_0=\mathcal{L}-\sum_i^N\mathcal{S}_i$ and $\mathcal{S}_i\hat{\rho}=\hat{d}_i\hat{\rho}\hat{d}_i^\dagger$ defines the collapse superoperator $\mathcal{S}_i$ \cite{carmichael} of the effective field $\hat{d}_i$ at the $i^\text{th}$ single-photon detector. For direct photon counting detection where vacuum fluctuations do not contribute, this effective field operator, which also accounts for photon losses, is described by the operators of the open quantum system and is referred to as the source field \cite{carmichael1989photoelectron,kiraz2004quantum,barrett2005efficient}. 

From equation (\ref{meqC}), the density matrix solution $\hat{\rho}(t)$ can be decomposed into a set of conditional states dependent on the cumulative detected photon count $n_i$ in the $i^\text{th}$ mode of $N$ modes using the Liouville-Neumann series \cite{zoller,carmichael}
\begin{equation}
    \hat{\rho}(t) = \sum_{i=1}^N\sum_{n_i=0}^\infty \hat{\rho}_{(n_1,n_2,\cdots,n_N)}(t) = \sum_{\mathbf{n}\in\mathbb{N}_0^N}\hat{\rho}_\mathbf{n}(t),
\end{equation}
where $\mathbf{n}=\sum_i^N n_i\mathbf{e}_i$, and $\mathbf{e}_i$ is the $i^\text{th}$ natural basis vector in the $N$-dimensional space of non-negative integers $\mathbb{N}_0^N$. The conditional state $\hat{\rho}_\mathbf{n}(t)$ is the unnormalized density matrix $\hat{\rho}_{\mathbf{n}}(t) = \mathcal{U}_\mathbf{n}(t,t_0)\hat{\rho}(t_0)$, where $\mathcal{U}_\mathbf{n}$ is the conditional propagation superoperator described recursively by
\begin{equation}
\label{conditionalpropagator}
    \mathcal{U}_\mathbf{n}(t,t_0) = \sum_{i=1}^N\int_{t_0}^t\mathcal{U}_0(t,t^\prime)\mathcal{S}_i\mathcal{U}_{\mathbf{n}-\mathbf{e}_i}(t^\prime,t_0)dt^\prime,
\end{equation}
and $\mathcal{U}_\mathbf{0}$ is the propagation superoperator of the equation $\dot{\rho} = \mathcal{L}_0\hat{\rho}(t)$. For convenience, we define $\mathcal{U}_\mathbf{n}=0$ if $\mathbf{n}\notin \mathbb{N}_0^N$. 

The conditional state $\hat{\rho}_\mathbf{n}(t)$ occurs with the probability $p_\mathbf{n}(t) = \text{Tr}[\hat{\rho}_\mathbf{n}(t)]$ where $\sum_{\mathbf{n}}p_\mathbf{n}=1$ and the resulting final state of the system at time $t$ is given by $\hat{\rho}_\mathbf{n}(t)/p_\mathbf{n}(t)$. This decomposition is an exact description of the original master equation dynamics because the total propagation superoperator of equation (\ref{meq}) is given by $\mathcal{U}(t,t_0) = \sum_{\mathbf{n}}\mathcal{U}_\mathbf{n}(t,t_0).$ As a consequence, this decomposition provides access to the state of the systems after post selecting based on the number of detected photons.

Since the total propagation superoperator $\mathcal{U}$ has the property that $\mathcal{U}(t,t_0) = \mathcal{U}(t,t^\prime)\mathcal{U}(t^\prime,t_0)$, we can also discuss conditional states for a window of time $T=t^\prime-t$ between an initial time $t_0$ and a final time $t_\text{f}$. The conditional propagator for this window is given by
\begin{equation}
\label{windowpropagator}
    \mathcal{W}_\mathbf{n}(t_\text{f},t^\prime,t,t_0) = \mathcal{U}(t_\text{f},t^\prime)\mathcal{U}_\mathbf{n}(t^\prime,t)\mathcal{U}(t,t_0),
\end{equation}
where $\mathcal{U}=\sum_{\mathbf{n}}\mathcal{W}_\mathbf{n}$. Later in this Methods section, we discuss how these conditional propagators can be used to define a gated photon counting measurement.

\subsection{Imperfections}
\label{subsec:imperfections}

We consider five main imperfections in the entanglement generation process: (1) decoherence, (2) spectral diffusion, (3) photon loss, (4) phase errors, and (5) dark counts. Solid-state systems may suffer from mechanisms that degrade the spin coherence and the coherence of emitted photons. These mechanisms are usually strongly dependent on temperature \cite{fu2009observation,plakhotnik2015electron,grange2017reducing}. These systems can also experience spectral diffusion, which can inhibit the indistinguishability of emitted photons \cite{loredo2016scalable,thoma2016exploring,reimer2016overcoming}. In addition, photon losses due to non-radiative pathways or collection/transmission inefficiency can affect the protocol figures of merit; and in some cases, protocols can moreover be susceptible to initialization and propagation phase errors. Finally, the detectors may have a non-negligible dark count rate \cite{hadfield2009single}.

\emph{Decoherence.---}For each protocol, we consider that a transition with decay rate $\gamma$ is subject to a pure dephasing rate $\gamma^\star$ that degrades the coherence of photons emitted by the system \cite{grange2015cavity}. This dephasing is due to fluctuations of the transition energy on a timescale much faster than its decay rate, and it affects the indistinguishability between photons emitted by the systems. We separate the decay rate of the transition into a radiative component $\gamma_\text{r}$ and a non-radiative component $\gamma_\text{nr}$ so that $\gamma=\gamma_\text{r}+\gamma_\text{nr}$. In addition, we consider that the spin qubits experience incoherent spin flip excitation (decay) at the rate $\gamma_\text{s}^+=1/T_1^+$ ($\gamma_\text{s}^-=1/T_1^-$) and a pure dephasing at the rate $\chi^\star=1/T_2^\star$ for a total spin decoherence rate of $1/T_2 = 1/T_2^\star+1/2T_1^++1/2T_1^-$.

\emph{Spectral diffusion.---}In contrast to pure dephasing, spectral diffusion is a fluctuation of the transition energy on a timescale much slower than its decay rate. In many solid-state systems, this fluctuation can shift the emitted photon frequency by more than its linewidth \cite{thoma2016exploring,reimer2016overcoming}. This degrades the mean wavepacket overlap of photons emitted by the same source at different times \cite{loredo2016scalable,thoma2016exploring,reimer2016overcoming}. Hence, this fluctuation also significantly degrades interference between fields from different sources. We account for spectral diffusion by averaging entanglement figures of merit over a Gaussian distribution \mbox{$f(\omega_k-\overline{\omega}_k,\delta_k) =(\delta_k\sqrt{2\pi})^{-1}e^{-(\omega_k-\overline{\omega}_k)^2/2\delta_k^2}$} for each emitter frequency $\omega_k$ with an average value of $\overline{\omega}_k$ and a spectral diffusion standard deviation $\delta_k$. For example, for two systems, the entanglement fidelity $F$ becomes \mbox{$\iint f(\omega_1-\overline{\omega}_1,\delta_1)f(\omega_2-\overline{\omega}_2,\delta_2)Fd\omega_1 d\omega_2$}.

\emph{Photon loss.---}Losses can occur due to non-radiative transitions at a rate $\gamma_\text{nr}$. We also quantify the imperfect collection fraction $\eta_\text{c}$ of emission and the fraction of photons transmitted to the detectors by $\eta_\text{t}$. In addition, we consider that each detector has a probability $\eta_\text{d}$ of detecting an incident photon. This detector inefficiency can be applied during the measurement step. However, the beam-splitter loss model used to describe detector inefficiency can be mapped to the identical model for transmission loss \cite{wein2016efficiency}. Thus, for convenience, we choose to simulate detector inefficiency as part of the conditional dynamics rather than the measurement itself. This allows us to use the total efficiency parameter $\eta=\eta_\text{c}\eta_\text{t}\eta_\text{d}\gamma_\text{r}/\gamma$.

\emph{Phase errors.---}Phase errors can arise when the individual quantum systems are locally initialized and read out using pulses from a source that does not maintain phase stability over the duration of the protocol. We account for this by considering an initial phase $\varphi$ when a quantum system is initialized in a superposition state. Phase errors can also arise when photons from each source do not accumulate the same propagation phase $\phi$ before interference. If these phases are unstable or left uncorrected, then they may degrade the entanglement fidelity. We account for phase errors by assuming that the phase fluctuates between entanglement generation attempts and then average the fidelity over a random phase with a Gaussian distribution.

\emph{Dark counts.---}A realistic detector may falsely indicate the arrival of a photon or detect a photon that did not originate from a desired emitter \cite{hadfield2009single}. In our study, we assume that each detector is gated for an interval $T_\text{d}$ that begins at time $t_\text{d}$ after the start of the protocol and ends at time $t_\text{d}^\prime=t_\text{d}+T_\text{d}$. We also assume that the dark counts are classical noise described by a Poisson distribution with a rate $\lambda$. Then for a given detector, the probability that $n$ dark counts have occurred during the gate duration $T_\text{d}$ is given by $\xi_n(T_\text{d},\lambda)=\lambda^nT_\text{d}^ne^{-\lambda T_\text{d}}/n!$.

\subsection{Measurement}\label{measurementsec}

For a gated detector operating at a distance $L_\text{d}$ from the emitter, the measurement depends on the state of the system at the retarded time $r(t) = t - L_\text{d}/c$, where $c$ is the transmission speed of light. The gated detector begins at time $t_\text{d}$ where $r(t_\text{d})\geq t_0$ and remains open for duration $T_\text{d}=t_\text{d}^\prime-t_\text{d}=r(t_\text{d}^\prime)-r(t_\text{d})$. The conditional state $\hat{\rho}_\mathbf{n}$ at time $t_\text{f}$ after a retarded detection window is
\begin{equation}
\label{equ:dmat_dis}
\begin{aligned}
    \hat{\rho}_\mathbf{n}(t_\text{f}) &= \mathcal{W}_\mathbf{n}(t_\text{f},r(t_\text{d}^\prime),r(t_\text{d}),t_0)\hat{\rho}(t_0),
\end{aligned}
\end{equation}
where $t_\text{f}-t_0=T_\text{d}+2L_\text{d}/c$ is the minimum protocol time after a two-way classical communication.

Let $\mathbb{M}$ be the space of all measurement outcomes. Then the system state after the outcome \mbox{$\mathbf{m}\in\mathbb{M}$} of the field state is communicated back to the system is
\begin{equation}
\label{eq:POVM}
\begin{aligned}
    \hat{\varrho}_\mathbf{m}(t_\text{f}) =\sum_{\mathbf{n}\in\mathbb{N}_0^N} P(\mathbf{m}|\mathbf{n})\hat{\rho}_\mathbf{n}(t_\text{f}),
\end{aligned}
\end{equation}
where $P(\mathbf{m}|\mathbf{n})$ is the probability for outcome $\mathbf{m}$ given state $\hat{\rho}_\mathbf{n}$. Note that we distinguish between the conditional state of the system $\hat{\rho}_\mathbf{n}$, where $\mathbf{n}$ denotes the true photon distribution, and the state after the measurement $\hat{\varrho}_\mathbf{m}$, where $\mathbf{m}$ includes imperfections in the measurement such as dark counts. Naturally, we also require that $\sum_{\mathbf{m}\in\mathbb{M}} P(\mathbf{m}|\mathbf{n}) = 1$ for all $\mathbf{n}\in\mathbb{N}_0^N$. 

This approach is analogous to applying a positive-operator valued measure (POVM) \cite{helstrom1969quantum}, with the exception that $\hat{\rho}_\mathbf{n}$ is not computed via a projective measurement onto the system space, but rather by a projective photon number measurement of the state of the field at the detector. In this sense, the conditional propagation superoperator for a detection window $\mathcal{W}_\mathbf{n}$ can be related to an effective projection superoperator $\mathcal{P}_\mathbf{n}$ by
\begin{equation}
\begin{aligned}
    \mathcal{P}_\mathbf{n}(t_\text{f},t_\text{d}^\prime,t_\text{d},t_0)\hat{\rho}(t_\text{f})\!=\!\hat{\rho}_\mathbf{n}(t_\text{f})
    \!=\!\mathcal{W}_\mathbf{n}(t_\text{f},t_\text{d}^\prime,t_\text{d},t_0)\hat{\rho}(t_0).\\
\end{aligned}
\end{equation}
This effective projection superoperator depends on the history of the system and the detection window times; however, it is complete:
\begin{equation}
\begin{aligned}
    \!\!\sum_{\mathbf{n}\in\mathbb{N}_0^N} \!\! \mathcal{P}_\mathbf{n}(t_\text{f},t_\text{d}^\prime,t_\text{d},t_0)\hat{\rho}(t_\text{f})&=\!\!\!    \sum_{\mathbf{n}\in\mathbb{N}_0^N}  \mathcal{W}_\mathbf{n}(t_\text{f},t_\text{d}^\prime,t_\text{d},t_0)\hat{\rho}(t_0)\\
    &=\mathcal{U}(t_\text{f},t_0)\hat{\rho}(t_0)=\hat{\rho}(t_\text{f}),
\end{aligned}
\end{equation}
hence $\sum_{\mathbf{n}}\mathcal{P}_\mathbf{n} = \mathcal{I}$ and so we have $\sum_{\mathbf{m},\mathbf{n}}P(\mathbf{m}|\mathbf{n})\mathcal{P}_\mathbf{n} = \mathcal{I}$. Thus $\mathcal{F}_\mathbf{m}=\sum_\mathbf{n}P(\mathbf{m}|\mathbf{n})\mathcal{P}_\mathbf{n}$ can be interpreted as the effective POVM element for the outcome $\mathbf{m}$ of the measured environment of the open quantum system and $\hat{\varrho}_\mathbf{m}=\mathcal{F}_\mathbf{m}\hat{\rho}$ is the unnormalized state after the measurement, which occurs with the probability $\eta_\mathbf{m}=\text{Tr}[\hat{\varrho}_\mathbf{m}]$.

By assuming that the detectors are identical and independent, we simplify the conditional probability to $P(\mathbf{m}|\mathbf{n}) \equiv \prod_i^N P_\text{d}(m_i|n_i)$. For fast photon-number-resolving detectors (PNRDs) that can count all photons arriving during the gate duration $T_\text{d}$, we have $\mathbb{M}=\mathbb{N}_0^N$. Then $P_\text{d}(m|n)$ is given by all possible combinations of dark counts such that $n$ can appear to be $m$:
\begin{equation}
\label{eq:PNRD}
    P_\text{PNRD}(m|n)=\sum_{k=0}^\infty\delta_{m,k+n}\xi_k(T_\text{d},\lambda),
\end{equation}
where $\delta$ is the Kronecker delta and $\xi_k$ characterizes the dark count distribution. On the other hand, for a bin detector (BD) that simply indicates the presence of one or more photons arriving during the gate duration, then $\mathbb{M}=\Sigma^N$ is the set of binary vectors of length $N$ and $P_\text{d}(m|n)$ is given by
\begin{equation}
\label{eq:BD}
\begin{aligned}
    \hspace{-2mm}P_\text{BD}(m|n) &= \sum_{q=0}^\infty\delta_{m,\text{sgn}(q)}P_\text{PNRD}(q|n)\\
    &=\!\delta_{m,\text{sgn}(n)}\xi_0(T_\text{d},\lambda)\!+\!\delta_{m,1}\!\left(1\!-\!\xi_0(T_\text{d},\lambda)\right),
\end{aligned}
\end{equation}
where $\text{sgn}:\mathbb{N}_0\to\Sigma$ is the signum function.

The PNRD and BD models are appropriate for many different detector types \cite{hadfield2009single,wein2016efficiency}. For example, single-photon avalanche photodiodes (APDs) and superconducting nanowire single-photon detectors (SNSPDs) \cite{natarajan2012superconducting} can be modeled by BDs while transition edge sensors (TESs) \cite{rosenberg2005noise,lita2008counting} are considered as PNRDs \cite{wein2016efficiency}.

\subsection{Figures of merit}

\emph{Entanglement generation.---}After the protocol, the unnormalized final state $\hat{\varrho}_\mathbf{m}$ conditioned on the measurement outcome $\mathbf{m}$ is associated with the outcome probability $\eta_\mathbf{m}=\text{Tr}[\hat{\varrho}_\mathbf{m}]$. For a given measurement outcome $\mathbf{m}$, we denote the expected final pure state as $\ket{\psi_\mathbf{m}}$. Then the fidelity associated with outcome $\mathbf{m}$ is $F_\mathbf{m}=\bra{\psi_\mathbf{m}}\hat{\varrho}_\mathbf{m}\ket{\psi_\mathbf{m}}/\eta_\mathbf{m}$. Let $\mathbb{A}\subset\mathbb{M}$ be the set of accepted measurement outcomes where $\mathbf{m}\in\mathbb{A}$ indicates the expected state $\ket{\psi_\mathbf{m}}$. Then the total entanglement generation efficiency is $\eta_\text{gen} = \sum_{\mathbf{m}\in\mathbb{A}}\eta_\mathbf{m}$ and the associated average entanglement generation fidelity weighted by outcome efficiency is $F_\text{gen} =\sum_{\mathbf{m}\in\mathbb{A}}\eta_\mathbf{m}F_\mathbf{m}/\eta_\text{gen}$.

We can also compute the concurrence of the system after projecting the total system onto a two-qubit subsystem. The entanglement concurrence \cite{wootters2001entanglement} is given by
\begin{equation}
    C_\mathbf{m} = \text{max}\left(0,\sqrt{\alpha_1}-\sqrt{\alpha_2}-\sqrt{\alpha_3}-\sqrt{\alpha_4}\right),
\end{equation}
where $\alpha_i\geq\alpha_{i+1}$ is the $i^\text{th}$ eigenvalue of $(\mathcal{I}_\text{s}\hat{\varrho}_\mathbf{m})(\mathcal{Y}_\text{s}\hat{\varrho}_\mathbf{m}^*)/\eta_\mathbf{m}^2$. Here, $\mathcal{I}_\text{s}\hat{\rho} = (\hat{I}_\text{s}\otimes\hat{I}_\text{s})\hat{\rho}(\hat{I}_\text{s}\otimes\hat{I}_\text{s})$ and $\hat{I}_\text{s}=\hat{\sigma}^\dagger\hat{\sigma}+\hat{\sigma}\hat{\sigma}^\dagger$, where $\hat{\sigma}$ is the spin qubit lowering operator; and $\mathcal{Y}_\text{s}\hat{\rho} =(\hat{\sigma}_y\otimes\hat{\sigma}_y)\hat{\rho}(\hat{\sigma}_y\otimes\hat{\sigma}_y)$, where $\hat{\sigma}_y = i(\hat{\sigma}-\hat{\sigma}^\dagger)$ is the Pauli $y$ operator. Then the weighted average concurrence is $C_\text{gen} = \sum_{\mathbf{m}\in\mathbb{A}}\eta_\mathbf{m}C_\mathbf{m}/\eta_\text{gen}$.

\emph{Individual quantum systems.---}Two common quantities used to characterize the performance of single-photon quantum emitters are the total emission brightness $\beta_k = \int_{t_\text{d}}^{t_\text{d}^\prime}\braket{\hat{a}_k^\dagger(t)\hat{a}_k(t)}dt$ and the mean wavepacket overlap
\begin{equation}
\label{mkdefinition}
    M_k = \frac{2}{\beta_k^2}\int_{t_\text{d}}^{t_\text{d}^\prime}\!\!\int_t^{t_\text{d}^\prime}\!|\!\braket{\hat{a}_k^\dagger(\tau)\hat{a}_k(t)}\!|^2d\tau dt,
\end{equation}
where $\hat{a}_k$ is the field operator of the collected mode \cite{kiraz2004quantum,grange2015cavity,wein2018feasibility,ghobadi2019progress}. The mean wavepacket overlap between photons from a single source is derived from the Hong-Ou-Mandel (HOM) interference \cite{hong1987measurement} visibility $V_\text{HOM}=1-2p_{11}$ where $p_{11}$ is the normalized probability for a coincident count. The more general mean wavepacket overlap expression for photons from two different sources can be derived using the methods of Ref. \cite{kiraz2004quantum} as outlined in the supplementary of Ref. \cite{ollivier2020}:
\begin{equation}
\label{m12definition}
\begin{aligned}
    M_{kl} =\!
    \frac{2}{\beta_k\beta_l}\!\int_{t_\text{d}}^{t_\text{d}^\prime}\!\!\int_t^{t_\text{d}^\prime}\!\!\text{Re}\!\left[\braket{\hat{a}_k^\dagger(\tau)\hat{a}_k(t)}^{\!*}\!\!\braket{\hat{a}_l^\dagger(\tau)\hat{a}_l(t)}\right]\!d\tau dt.
\end{aligned}
\end{equation}

For finite excitation pulses, the emitter may also have a non-negligible chance for re-excitation. This is characterized by the integrated intensity correlation \cite{ghobadi2019progress}
\begin{equation}
    g^{(2)}_k \!=\! \frac{2}{\beta_k^2}\int_{t_\text{d}}^{t_\text{d}^\prime}\!\!\int_t^{t_\text{d}^\prime}\!\!\braket{\hat{a}_k^\dagger(t)\hat{a}_k^\dagger(\tau)\hat{a}_k(\tau)\hat{a}_k(t)}d\tau dt.
\end{equation}
However, for all the cases presented in this work, we assume that excitation pulses are fast enough compared to the timescale of other system dynamics so that $g^{(2)}\simeq 0$. Although it is beyond the scope of this paper, future work should address the limits to entanglement generation fidelity due to non-zero $g^{(2)}$, as this is one factor currently limiting the interference visibility for resonantly excited single-photon sources \cite{bernien2013heralded,ollivier2020}.

\subsection{Three-level systems}
\label{subsec:systems}

The Lindblad master equation for $K$ independent three-level systems (see figure \ref{fig:Lsystem}a) in the Hilbert space $\mathbb{H}$ is $\dot{\rho}=\mathcal{L}\hat{\rho}$ where $\mathcal{L}\in(\mathbb{H}\otimes\mathbb{H})^{\otimes K}$ and $\mathcal{L}=\sum_k^K\mathcal{L}_k^{(k)}$ is the shorthand summation in the tensor space of independent superoperators $\mathcal{L}_k\in\mathbb{H}\otimes\mathbb{H}$ given by
\begin{equation}
\begin{aligned}
    \mathcal{L}_k =& -i\mathcal{H}_k+\sum_j\gamma_{j_k}^-\mathcal{D}(\hat{\sigma}_{j})+\gamma_{j_k}^+\mathcal{D}(\hat{\sigma}_{j}^\dagger)\\
    &+2\gamma_k^\star\mathcal{D}(\hat{\sigma}_{\uparrow}^\dagger\hat{\sigma}_{\uparrow})+ \frac{\chi_k^\star}{2}\mathcal{D}(\hat{\sigma}_z),
\end{aligned}
\end{equation}
where we take $\hbar=1$, $\mathcal{H}\hat{\rho}=[\hat{H},\hat{\rho}]$, and $\mathcal{D}(\hat{\sigma})\hat{\rho} = \hat{\sigma}\hat{\rho}\hat{\sigma}^\dagger - \{\hat{\sigma}^\dagger\hat{\sigma},\hat{\rho}\}/2$. The system operators are defined $\hat{\sigma}_{\uparrow}=\ket{\uparrow}\!\bra{e}$, $\hat{\sigma}_{\downarrow}=\ket{\downarrow}\!\bra{e}$, $\hat{\sigma}_\text{s}=\ket{\uparrow}\!\bra{\downarrow}$, and $\hat{\sigma}_{z}=\ket{\uparrow}\!\bra{\uparrow}-\ket{\downarrow}\!\bra{\downarrow}$. The rate $\gamma_{j_k}^-$ ($\gamma_{j_k}^+$) is the total incoherent decay (excitation) rate across the transition associated with $\hat{\sigma}_j$ where $j\in\{\uparrow,\downarrow,\text{s}\}$, $\gamma_k^\star$ is the optical pure dephasing rate, and $\chi^\star_k$ is the spin pure dephasing rate. The three-level system Hamiltonian is $\hat{H}_k = \omega_{\uparrow_k}\hat{\sigma}_{\uparrow}^\dagger \hat{\sigma}_{\uparrow} + \omega_{\text{s}_k}\hat{\sigma}_\text{s}^\dagger\hat{\sigma}_\text{s}\!$ where $\omega_{\uparrow_k}$ is the separation between $\ket{\uparrow}$ and $\ket{e}$, $\omega_{\text{s}_k}$ is the separation between $\ket{\uparrow}$ and $\ket{\downarrow}$.

\subsection{Computational techniques}

To solve the system dynamics, we make use of the Fock-Liouville space representation of the master equation, $d\ket{\rho}\!\rangle/dt=\tilde{\mathcal{L}}\ket{\rho}\!\rangle$, where $\ket{\rho}\!\rangle$ is the vector representation of the density operator $\hat{\rho}$ and $\tilde{\mathcal{L}}$ is the matrix representation of the superoperator $\mathcal{L}$ \cite{manzano2020short}. For a superoperator of the form $\mathcal{A}\hat{\rho}=\sum_i\hat{A}_i\hat{\rho}\hat{B}_i$, where $\hat{A}_i$ and $\hat{B}_i$ are operators acting on the total Hilbert space $\mathbb{H}^K$, the matrix representation can be obtained using the relation $\tilde{\mathcal{A}} = \sum_i\hat{A}_i\otimes\hat{B}_i^\text{T}$, where $\hat{B}_i^\text{T}$ is transpose of $\hat{B}_i$.

The Fock-Liouville representation can also be used to solve the evolution of the conditional state $\hat{\rho}_\mathbf{0}$, the only difference being that $\mathcal{L}_\mathbf{0}$ does not preserve the trace of $\hat{\rho}$. If $\mathcal{L}_\mathbf{0}$ is independent of time, then the corresponding propagation superoperator can be computed in the matrix representation using standard techniques to solve $\tilde{\mathcal{U}}_\mathbf{0}(t_\text{f},t_0)=e^{(t_\text{f}-t_0)\tilde{\mathcal{L}}_\mathbf{0}}$ by diagonalization. The remaining conditional propagators $\tilde{\mathcal{U}}_\mathbf{n}$ for $\mathbf{n}\neq \mathbf{0}$ are then computed by recursive application of equation (\ref{conditionalpropagator}).

In some cases, $\tilde{\mathcal{L}}_\mathbf{0}$ can be analytically diagonalized, providing analytic solutions for the conditional propagation superoperators. Otherwise, $\tilde{\mathcal{L}}_\mathbf{0}$ can be numerically diagonalized for a fixed set of parameters resulting in $\tilde{\mathcal{U}}(t_\text{f},t_0)$ that is still analytic with respect to time. For smaller systems, such as those presented in this work, this approach can drastically decrease the time needed to compute the time dynamics by allowing equation (\ref{conditionalpropagator}) to be analytically solved for an arbitrary detection interval. For larger or time-dependent systems, numerical integration methods such as Runge-Kutta could also be used.

\section{Protocols}
\label{sec:protocols}

We now apply the method outlined in the previous section to analyze three different entanglement generation protocols. These three protocols rely on fast, pulsed, resonant excitation of the quantum systems. Hence, the properties of the emitted single photons are dominated by the properties of the quantum system rather than by the properties of the excitation pulses. Under this assumption, to simplify the analysis we consider all preparation and excitation pulses to be instantaneous perfect operations. However, we emphasize that the photon count decomposition can be applied to any Markovian master equation, including those with driving Hamiltonians and complicated time-dependent parameters.

\subsection{Spin-photon number entanglement (protocol $\mathsf{N}$)}
\label{subsec:phtnumber}

\begin{figure}
\includegraphics[width=8.5cm]{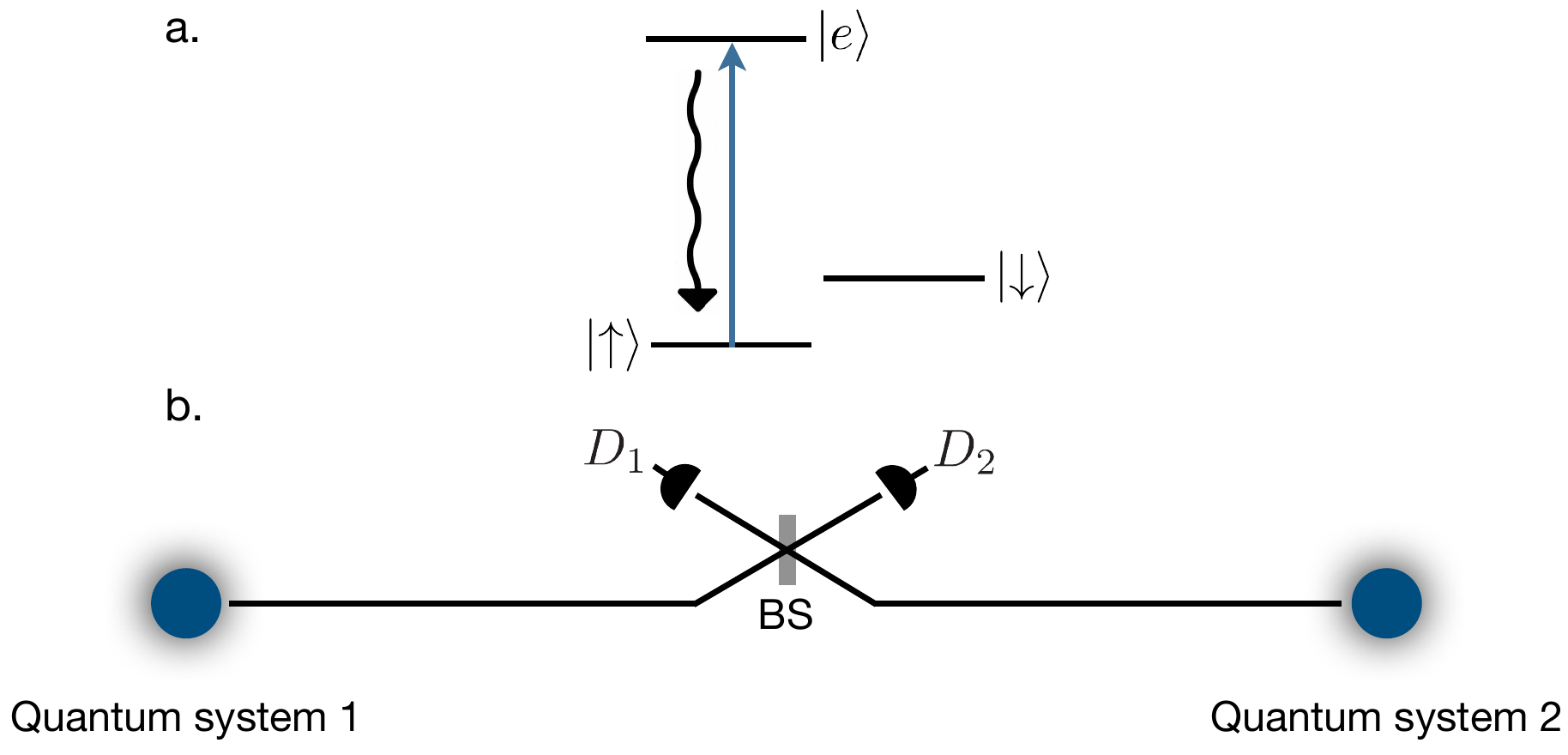}
\caption{\textbf{Entangling pulsed L systems.} (a) An L-type system as used in protocols $\mathsf{N}$ and $\mathsf{T}$; a ground state doublet $\ket{\uparrow}$ and $\ket{\downarrow}$ with one optically excited state $\ket{e}$ that decays back to its initial ground state $\ket{\uparrow}$. (b) An diagram illustrating the fields from each quantum system interfering at a central beam splitter (BS) that has its output ports monitored by single-photon detectors $\text{D}_1$ and $\text{D}_2$.}\label{fig:Lsystem}
\end{figure}

Consider the scheme where two spatially separated L-type systems are entangled by heralding a single photon emission after erasing the which-path information using a beam splitter (see figure \ref{fig:Lsystem}a). This scheme is similar to the scheme used in the DLCZ repeater protocol to generate entanglement between spatially separated quantum memories \cite{duan2001long}. However, by using fast resonant pulses, the quantum system requires only one optical transition. This scheme generates spin-spin entanglement by using spin-photon number entanglement \cite{rozpkedek2019near}. For this reason, we will denote it as protocol $\mathsf{N}$. For this protocol, we assume that the excited state $\ket{e}$ can only decay to spin state $\ket{\uparrow}$. That is, we assume that $\gamma_{\downarrow_k}^-\ll \gamma_{\uparrow_k}^-=\gamma_k$, where $k\in\{1,2\}$ indexes the system.

\emph{Protocol description.---}Each system is first prepared in the state $\ket{\downarrow}$. Then a microwave pulse resonant with the $\ket{\downarrow}\longleftrightarrow\ket{\uparrow}$ transition with a pulse area of $2\vartheta$ and phase $\varphi_k$ brings the spin qubit to the state $\cos(\vartheta)\ket{\downarrow}+\sin(\vartheta)e^{i\varphi_k}\ket{\uparrow}$. After an optical $\pi$-pulse is applied to excite $\ket{\uparrow}$, each system is left in a superposition of ground and excited states. The excited state then decays back to $\ket{\uparrow}$ and the system emits a photon with a probability $\sin^2(\vartheta)$. By perfectly interfering the fields from two quantum systems (see figure \ref{fig:Lsystem}b), the which-path information is erased and a single detection event will herald one of the Bell states $\ket{\psi^\pm}=(\ket{\uparrow\downarrow}\pm\ket{\downarrow\uparrow})/\sqrt{2}$. 

To show protocol $\mathsf{N}$ in detail, consider the simpler case where $\vartheta=\pi/4$ and $\varphi_k=0$. Then the total state of the quantum systems before decay is
\begin{equation}
\label{eq:singlephoton_initialstate}
    \ket{\psi(t_0)}=\frac{1}{2}(\ket{\downarrow}+\ket{e})_1\otimes(\ket{\downarrow}+\ket{e})_2.
\end{equation}
After decay, $\ket{e}\rightarrow\ket{\uparrow}$, each system is in a spin-photon number entangled state $\ket{\psi}_k=(\ket{\downarrow}\ket{0}+\ket{\uparrow}\ket{1})_k/\sqrt{2}$, where $\ket{0}$ is the vacuum state and $\ket{1}$ is the single photon state of emission mode. After interfering photons $\ket{1}_1$ and $\ket{1}_2$ at a beam splitter, the state before detection is
\begin{equation}
\label{eq:singlephoton_beamsplitter}
    \frac{1}{2}\left(\ket{\downarrow\downarrow}\ket{ 00}+\ket{\psi^+}\ket{01}+\ket{\psi^-}\ket{10}+\ket{\uparrow\uparrow}\ket{\psi^-_{2002}}
\right)_\text{s,p},\\
\end{equation}
where $\ket{n_1n_2}_\text{p}$ is the state with $n_1$ ($n_2$) photons in the mode of detector $\text{D}_1$ ($\text{D}_2$), \mbox{$\ket{\psi^\pm}_\text{s}\!=\!\left(\ket{\uparrow\downarrow}\pm\ket{\downarrow\uparrow}\right)/\sqrt{2}$} are spin Bell states, and $\ket{\psi^-_{2002}}_\text{p}=(\ket{20}-\ket{02})/\sqrt{2}$ is a two-photon NOON state. Hence, a single photon at either detector heralds a maximally entangled spin state with a phase determined by which detector received the photon.

The maximum efficiency of the above scheme is 50\%, which is the Bell analyzer efficiency of a single beam splitter \cite{calsamiglia2001maximum,wein2016efficiency}. However, any amount of photon loss will cause infidelity due to states $\ket{20}$ and $\ket{02}$ contributing to single-photon measurement outcomes. If $\vartheta$ is small enough, then the probability for both quantum systems to emit photons becomes much less than the probability that only one system emits a photon. Thus, to combat infidelity due to multi-photon events, the parameter $\vartheta$ can be reduced to improve fidelity at the cost of efficiency \cite{rozpkedek2019near}. This trade-off also improves the protocol fidelity for BD type detectors.

\emph{Conditional states.---}In the far field approximation, the source field component collected from a quantum emitter dipole is described by $\hat{a}_k=\hat{\sigma}_{\uparrow_k}\sqrt{\eta_{\text{c}_k}\gamma_{\text{r}_k}}$ \cite{carmichael,kiraz2004quantum,fischer2016dynamical}. After transmission losses and a propagation phase we have $\hat{a}_k\rightarrow\hat{a}_k\sqrt{\eta_{\text{t}_k}}e^{-i\phi_k}$ where $\phi_k=L_k\omega_{\uparrow_k}/v_\text{p}$, $L_k$ is the propagation distance, and $v_\text{p}$ is the phase velocity. Then the fields are interfered at a beam splitter so that the fields $\hat{d}_1$ and $\hat{d}_2$ at detectors D$_1$ and D$_2$, respectively, are
\begin{equation}
\label{beamsplittereq}
    \begin{pmatrix}
    \hat{d}_1\\
    \hat{d}_2\\
    \end{pmatrix}
    =
    \hat{R}(\theta)    \begin{pmatrix}
    \hat{a}_1\sqrt{\eta_{\text{t}_1}}e^{-i\phi_1}\\
    \hat{a}_2\sqrt{\eta_{\text{t}_2}}e^{-i\phi_2}\\
    \end{pmatrix},
\end{equation}
where $\hat{R}(\theta)$ is the $2\times 2$ rotation unitary matrix. Using a beam-splitter model for detector inefficiency, the effective detected field is $\hat{d}_i\rightarrow\hat{d}_i\sqrt{\eta_{\text{d}_i}}$. In this case, the collapse superoperators for the effective detected field at the $i^\text{th}$ detector are $\mathcal{S}_i\hat{\rho} = \eta_{\text{d}_i}\hat{d}_i\hat{\rho}\hat{d}_i^\dagger$, where to simplify calculations we will take $\eta_{\text{d}_i}=\eta_\text{d}$ for each detector.

If we assume that the pulse Rabi frequency is much faster than the rates of dissipation and decoherence, and that the excitation pulses are resonant with the emitters, then we can consider the initial state to be approximated by $\hat{\rho}(t_0)=\ket{\psi(t_0)}\!\bra{\psi(t_0)}_1\otimes\ket{\psi(t_0)}\!\bra{\psi(t_0)}_2$ where $\ket{\psi(t_0)}_k=\cos(\vartheta)\ket{\downarrow}+\sin(\vartheta)e^{i\varphi_k}\ket{e}$. Under these conditions, the set of conditional states can also be truncated to those $\mathbf{n}$ such that $n_1+n_2\leq 2$ as a consequence of each emitter only being able to emit up to a single photon.

The successful conditional states are associated with the single-photon detection conditions $\mathbf{n}=$ $(1,0)$, and $(0,1)$, which are given by their corresponding conditional propagators $\mathcal{W}_\mathbf{n}$. For convenience, we notate these vectors by 10, and 01, respectively. For example, the outcome associated with $\ket{\psi^+}$ is $\hat{\rho}_{01}(t_\text{f})=\mathcal{W}_{01}(t_\text{f},t_\text{d}^\prime,t_\text{d},t_0)\hat{\rho}(t_0)$ where $\mathcal{W}_{01}(t_\text{f},t_\text{d}^\prime,t_\text{d},t_0)=\mathcal{U}(t_\text{f},t_\text{d}^\prime)\mathcal{U}_{01}(t_\text{d}^\prime,t_\text{d})\mathcal{U}(t_\text{d},t_0)$ and
\begin{equation}
\label{singlephotonconditionalstates}
\begin{aligned}
    \mathcal{U}_{01}(t_\text{d}^\prime,t_\text{d})
    = \int_{t_\text{d}}^{t_\text{d}^\prime}\mathcal{U}_\mathbf{0}(t_\text{d}^\prime,t)\mathcal{S}_2\mathcal{U}_\mathbf{0}(t,t_\text{d})dt.
\end{aligned}
\end{equation}
Likewise, the outcome $\hat{\rho}_{10}$ associated with $\ket{\psi^-}$ is given by $\mathcal{W}_{10}$, which is determined using equation (\ref{singlephotonconditionalstates}) but with $\mathcal{S}_1$ in place of $\mathcal{S}_2$.
The states corresponding to the remaining relevant conditions $\mathbf{n}=$ (2,0), (1,1), and (0,2) are similarly obtained from equations (\ref{conditionalpropagator}) and (\ref{windowpropagator}).

\begin{figure}
    \centering
    \includegraphics[scale=0.5]{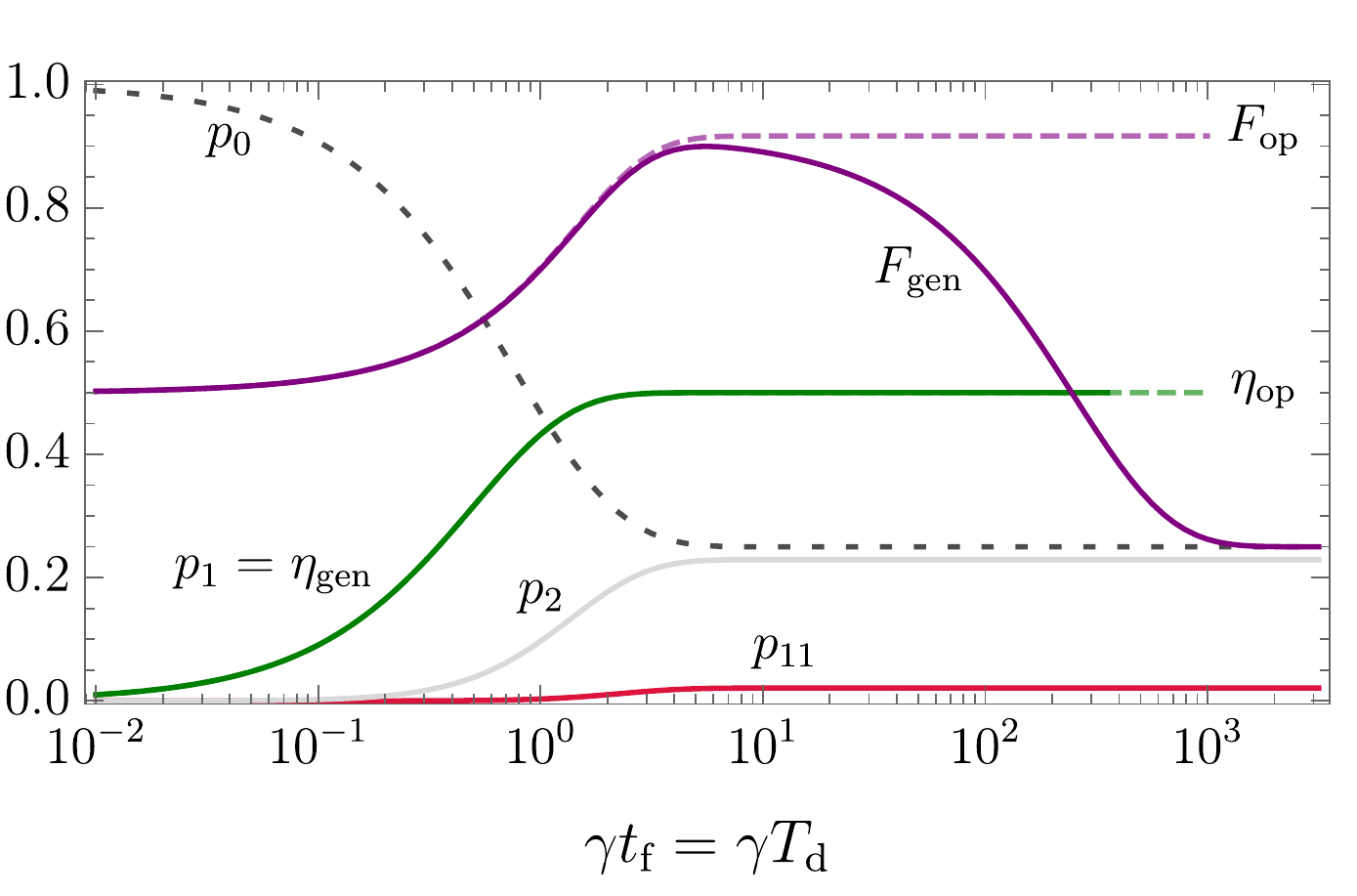}
    \caption{\textbf{Protocol $\mathsf{N}$---time dynamics.} Entanglement generation fidelity $F_\text{gen}$ and efficiency $\eta_\text{gen}$ as a function of protocol time $t_\text{f}$ for an initial state $\hat{\rho}(t_0)=\ket{\psi(t_0)}\!\bra{\psi(t_0)}$ where $\ket{\psi(t_0)}=(1/2)(\ket{\downarrow}+\ket{e})^{\otimes 2}$ and where there is no loss and noiseless local photon-number resolving detectors. The asymptotic dashed lines indicate the limits on fidelity $F_\text{op}$ and efficiency $\eta_\text{op}$ reached when spin decoherence is neglected and when the detection window encompasses the entire photon lifetime. The detection window is set to be equal to the entire protocol duration: $t_\text{d}=t_0=0$ and $t_\text{d}^\prime=t_\text{f}=T_\text{d}$. The gray lines show the probabilities for no photon emission from either system $p_0=\text{Tr}[\hat{\rho}_\mathbf{0}]$ and for photon bunching $p_2=p_{20}+p_{02}=\text{Tr}[\hat{\rho}_{20}+\hat{\rho}_{02}]$. The red line indicates coincident counts $p_{11}=\text{Tr}[\hat{\rho}_{11}]$ caused by imperfect HOM interference. Parameters chosen: $\gamma_k=\gamma$, $\gamma_k^\star=0.1\gamma$, and $\gamma_{\text{s}_k}^\pm=\chi_k^\star=0.001\gamma$ for $k\in\{1,2\}$.}
    \label{fig:onephotonTimeDynamics}
\end{figure}

\emph{Measurement duration.---}The spin entanglement is generated at the moment a single photon from one of the emitters is detected by one of the two detectors. However, a subsequent detection of another photon will destroy this entanglement \cite{martin2019single}. As a consequence, unless the probability for two-photon events is very small, the detection duration must be long enough to ensure that only one photon was emitted. Hence the fidelity can be very low for small $T_\text{d}$. On the other hand, for a detection window much longer than the lifetime, the fidelity becomes limited by spin decoherence processes. Figure \ref{fig:onephotonTimeDynamics} shows the entanglement generation fidelity and efficiency as the detection window duration is increased, illustrating the peak in fidelity when the duration is on the order of the lifetime. To show this qualitative behaviour, we have chosen parameters in the regime $\gamma>\gamma^\star\gg\chi^\star$ to represent a solid-state system that could potentially serve as a quantum communication node. The two-photon probabilities for photon bunching and coincident counts are also illustrated. Note that the coincident counts $p_{11}$ are nonzero due to the optical pure dephasing that degrades the HOM interference \cite{hong1987measurement} between photons from different sources.

Under the condition where the systems experience negligible spin decoherence on the timescale of the lifetime of the emitter, we can analytically solve the conditional states as a function of detection window duration. This can then be used to estimate simple figures of merit for the quality of entanglement based only on the optical properties of the emitters. These limits are illustrated by the asymptotes of the dashed lines in figure \ref{fig:onephotonTimeDynamics}. The analytic solutions can also then be used to estimate the fidelity under the effects of additional imperfections such as spectral diffusion and noisy detectors using the methods outlined in Sec. \ref{subsec:imperfections} and \ref{measurementsec}.

Suppose that the detection window begins at $t_\text{d}$ such that $r(t_\text{d})=t_0$ and ends at $t_\text{d}^\prime$ such that $r(t_\text{d}^\prime)=T_\text{d}+t_0$. Using the appropriate conditional propagators $\mathcal{W}_\mathbf{n}$, we compute the final (unnormalized) spin-spin conditional states in the rotating frame of the spin qubits after time $t_\text{f}\gg1/\gamma_k$. In this limit of time, neither quantum system remains in $\ket{e}$.

The conditional spin state of the quantum systems given that both detector modes do not contain a photon from an emitter is
\begin{equation}
\begin{aligned}
    &\hat{\rho}_\mathbf{0} = \frac{1}{4}\sin^2(2\vartheta)\left((1-\overline{\beta}_1)\ket{\uparrow\downarrow}\!\bra{\uparrow\downarrow}+(1-\overline{\beta}_2)\ket{\downarrow\uparrow}\!\bra{\downarrow\uparrow}\right)\\ &+(1-\overline{\beta}_1)(1-\overline{\beta}_2)\sin^4(\vartheta)\ket{\uparrow\uparrow}\!\bra{\uparrow\uparrow}+\cos^4(\vartheta)\ket{\downarrow\downarrow}\!\bra{\downarrow\downarrow},\\
\end{aligned}
\end{equation}
where $\overline{\beta}_k=\eta_\text{d}\beta_k/\gamma_k=\eta_k\left(1-e^{-T_\text{d}\gamma_k}\right)$ is the rate-normalized brightness and $\eta_k=\eta_\text{d}\eta_{\text{t}_k}\eta_{\text{c}_k}\gamma_{\text{r}_k}/\gamma_k$ is the total single-photon efficiency of each quantum system.

The single-photon conditioned states of the quantum system are
\begin{equation}
\begin{aligned}
    \hat{\rho}_\text{s}^\pm= &\rho_{\uparrow\uparrow}^\pm\ket{\uparrow\uparrow}\!\bra{\uparrow\uparrow}+\rho_{\uparrow\downarrow}^\pm\ket{\uparrow\downarrow}\!\bra{\uparrow\downarrow}+\rho_{\downarrow\uparrow}^\pm\ket{\downarrow\uparrow}\!\bra{\downarrow\uparrow}\\
    &+\left(\rho_\text{c}^\pm\ket{\downarrow\uparrow}\!\bra{\uparrow\downarrow}+\text{H.c.}\right),
\end{aligned}
\end{equation}
where
\begin{equation}
\begin{aligned}
\label{P1condsol}
    \rho_{\uparrow\uparrow}^\pm &= \frac{1}{2}\!\left(\overline{\beta}_1\!+\!\overline{\beta}_2\!-\!2\overline{\beta}_1\overline{\beta}_2\pm(\overline{\beta}_1\!-\!\overline{\beta}_2)\cos(2\theta)\right)\!\sin^4\!\left(\vartheta\right)\\
    \rho_{\downarrow\uparrow}^\pm &=\frac{\overline{\beta}_1}{8}(1\pm\cos(2\theta))\sin^2(2\vartheta)\\
    \rho_{\uparrow\downarrow}^\pm &= \frac{\overline{\beta}_2}{8}(1\mp\cos(2\theta))\sin^2(2\vartheta)\\
    \rho_\text{c}^\pm \!&= \pm\frac{\tilde{C}(T_\text{d})}{8}\sqrt{\eta_1\eta_2}\sin(2\theta)\sin^2(2\vartheta)\\
    \tilde{C}(T_\text{d})&=\frac{2\sqrt{\gamma_
1\gamma_2}}{\Gamma_1+\Gamma_2+2i\Delta}\!\left(1-e^{-\frac{1}{2}T_\text{d}(\Gamma_1+\Gamma_2+2i\Delta)}\right)\!e^{i(\varphi+\phi)},
\end{aligned}
\end{equation}
and where $\Gamma_k = \gamma_k+2\gamma_k^\star$ is the FWHM of the emission ZPL for system $k$, $\Delta=\omega_{\uparrow_1}-\omega_{\uparrow_2}$ is the spectral detuning, $\varphi=\varphi_1-\varphi_2$ is the relative initialization phase, and $\phi=\phi_1-\phi_2$ is the relative propagation phase. The sign of $\hat{\rho}_\text{s}^\pm$ is given by which detector received the photon: $\hat{\rho}_{10}=\hat{\rho}_\text{s}^-$ and $\hat{\rho}_{01}=\hat{\rho}_\text{s}^+$. Note that $\rho_{\downarrow\downarrow}=0$ because after either system emits one photon, the system is guaranteed to not be in $\ket{\downarrow\downarrow}$.

The individual two-photon conditioned states $\hat{\rho}_{20}$, $\hat{\rho}_{11}$, and $\hat{\rho}_{02}$ are all proportional to $\ket{\uparrow\uparrow}\!\bra{\uparrow\uparrow}$, as expected. However, their trace has a complicated dependency on $T_\text{d}$ due to the HOM effect. Regardless, their sum can be easily simplified to the intuitive result
\begin{equation}
    \hat{\rho}_{20}+\hat{\rho}_{11}+\hat{\rho}_{02}=\overline{\beta}_1\overline{\beta}_2\sin^4(\vartheta)\ket{\uparrow\uparrow}\!\bra{\uparrow\uparrow}.
\end{equation}
Using all these conditional states, we can also verify that
\begin{equation}
\begin{aligned}
    &\hat{\rho}_\mathbf{0}+\hat{\rho}_\text{s}^++\hat{\rho}_\text{s}^-+\hat{\rho}_{20}+\hat{\rho}_{11}+\hat{\rho}_{02}=\sin^4(\vartheta)\ket{\uparrow\uparrow}\!\bra{\uparrow\uparrow}\\&+\cos^4(\vartheta)\ket{\downarrow\downarrow}\!\bra{\downarrow\downarrow}+\frac{1}{4}\sin^2(2\vartheta)\left(\ket{\uparrow\downarrow}\!\bra{\uparrow\downarrow}+\ket{\downarrow\uparrow}\!\bra{\downarrow\uparrow}\right)\\
\end{aligned}
\end{equation}
is the solution $\hat{\rho}(t)$ of the total master equation in the limit $t\gg1/\gamma_k$ and when spin decoherence is neglected.

The states $\hat{\rho}_\text{s}^\pm$ correspond to the expected Bell states $\ket{\psi^\pm}$ and so the average entanglement fidelity is
\begin{equation}
    F_\text{gen} = \frac{1}{\eta_\text{gen}}\left(\bra{\psi^+}\hat{\varrho}_\text{s}^+\ket{\psi^+}+\bra{\psi^-}\hat{\varrho}_\text{s}^-\ket{\psi^-}\right),
\end{equation}
where $\eta_\text{gen} = \text{Tr}\left[\hat{\varrho}_\text{s}^+\right]+\text{Tr}\left[\hat{\varrho}_\text{s}^-\right]$ is the entanglement generation efficiency and $\hat{\varrho}_\text{s}^\pm$ is the state after measurement computed from the conditional state $\hat{\rho}_\text{s}^\pm$.

\emph{Optical limits.---}Suppose that the interference is balanced so that $\theta=\pi/4$ and $\eta_1=\eta_2=\eta$. Also, suppose that the protocol is phase corrected so that $\varphi+\phi=0$ (see section \ref{phaseerrors} for a discussion on phase errors). Then in the limit that $T_\text{d}\gg 1/\gamma_k$ we have $\overline{\beta}_1\rightarrow \eta$, $\overline{\beta}_2\rightarrow \eta$, and
\begin{equation}
\label{ctilde}
    \tilde{C}\rightarrow\frac{2\sqrt{\gamma_1\gamma_2}}{\Gamma_1+\Gamma_2+2i\Delta}.
\end{equation}
If we also assume that the measurement is performed by ideal noiseless PNRDs, then $\hat{\varrho}_\text{s}^\pm=\hat{\rho}_\text{s}^\pm$ and under these conditions---which we refer to as the optical limit---the corresponding entanglement generation fidelity $F_\text{op}$ gives an estimate of the fidelity determined only by the optical properties of the emitters. In principle, this bound could be exceeded using spectral or temporal post selection of photons, consequently sacrificing efficiency.

In the optical limit, the fidelity for protocol $\mathsf{N}$ becomes 
\begin{equation}
\label{Nfidelity}
\begin{aligned}
     F_\text{op}&=\frac{1}{2}\left(1+\text{Re}(\tilde{C})\right)F_\eta(\vartheta)\\
\end{aligned}
\end{equation}
with concurrence \mbox{$C_\text{op}=|\tilde{C}|F_\eta(\vartheta)$}, where the loss compensation factor is \mbox{$F_\eta(\vartheta)=\cos^2(\vartheta)/(1\!-\!\eta\sin^2(\vartheta))$}. The efficiency becomes \mbox{$\eta_\text{op}=(\eta/2)\sin^2(2\vartheta)/F_\eta(\vartheta)$} and the two-photon conditioned states reduce to
\begin{equation}
\label{2photoncondstates}
\begin{aligned}
    \hat{\rho}_{20} &= \hat{\rho}_{20} = \frac{1}{4}\left(1+M_{12}\right)\eta^2\sin^4(\vartheta)\ket{\uparrow\uparrow}\!\bra{\uparrow\uparrow}\\
    \hat{\rho}_{11} &= \frac{1}{2}\left(1-M_{12}\right)\eta^2\sin^4(\vartheta)\ket{\uparrow\uparrow}\!\bra{\uparrow\uparrow},
\end{aligned}
 \end{equation}
where
\begin{equation}
\label{meanwavepacketoverlap}
    M_{12} = M_\gamma\frac{(\Gamma_1+\Gamma_2)(\gamma_1+\gamma_2)}{(\Gamma_1+\Gamma_2)^2+4\Delta^2}\leq \sqrt{M_1M_2},
\end{equation}
is the mean wavepacket overlap, $M_k=\gamma_k/\Gamma_k$ is the individual system indistinguishability from equation (\ref{mkdefinition}), and $M_\gamma=4\gamma_1\gamma_2/(\gamma_1+\gamma_2)^2\geq M_{12}$ quantifies the temporal profile mismatch. We emphasize that equation (\ref{2photoncondstates}) and $M_{12}$ in equation (\ref{meanwavepacketoverlap}) were solved using the methods of Sec. \ref{subsec:evolution} and not using equation (\ref{m12definition}). However, we have verified that solving equation (\ref{m12definition}) indeed gives the same result as equation (\ref{meanwavepacketoverlap}), which confirms that the photon statistics of the HOM interference are independent of whether the calculation is performed from the perspective of the emitter or the field.

For a given $F_\eta(\vartheta)$, the fidelity and concurrence are limited by the spectral and temporal properties of the individual emitters. In particular, we can identify that $C_\text{op}^2\leq M_{12}F^2_\eta(\vartheta)$. Hence for protocol $\mathsf{N}$, the square root of the mean wavepacket overlap $M_{12}$ gives an upper bound on the entanglement generation concurrence, which itself can be used to determine an upper bound on the entanglement generation fidelity by $F_\text{op} \leq\left(1+\sqrt{M_{12}}\right)F_\eta(\vartheta)$. On the other hand, we have that $\text{Re}(\tilde{C})=M_{12}/\sqrt{M_\gamma}\geq M_{12}$. Hence the optical limit of fidelity $F_\text{op}$ for protocol $\mathsf{N}$ is bounded by
\begin{equation}
    \frac{1}{2}\left(1+M_{12}\right)\leq\frac{F_\text{op}}{F_\eta(\vartheta)}\leq\frac{1}{2}\left(1+\sqrt{M_{12}}\right).
\end{equation}

\emph{Detector noise and number resolution.---}The entangled spin-spin state after a single-photon measurement by a PNRD with non-negligible noise is $\hat{\varrho}_\text{s}^\pm = \xi_0^2\hat{\rho}_\text{s}^\pm+ \xi_0\xi_1\hat{\rho}_\mathbf{0}$, where $\xi_n(T_\text{d},\lambda)$ is the probability to have $n$ dark counts within the detection window $T_\text{d}$. For $T_\text{d}\gg1/\gamma_k$, we can write
\begin{equation}
\label{fgennoise}
    F_\text{gen} = \frac{1}{\eta_\text{gen}}\left(\xi_0^2F_\text{op}\eta_\text{op}+\frac{1}{2}\xi_0\xi_1(1-\eta)\sin^2(2\vartheta)\right),
    \vspace{-1mm}
\end{equation}
where
\vspace{-1mm}
\begin{equation}
    \eta_\text{gen} = \xi_0^2\eta_\text{op} + 2\xi_0\xi_1\left(1-\eta\sin^2(\vartheta)\right)^2
\end{equation}
is the total efficiency. For a measurement by a BD, the state after heralding is given by
\begin{equation}
\label{bdnoise}
\begin{aligned}
    \hat{\varrho}_\text{s}^-&=\xi_0(\hat{\rho}_\text{s}^-+\hat{\rho}_{20}) + \xi_0(1-\xi_0)\hat{\rho}_\mathbf{0}\\
    \hat{\varrho}_\text{s}^+&=\xi_0(\hat{\rho}_\text{s}^++\hat{\rho}_{02}) + \xi_0(1-\xi_0)\hat{\rho}_\mathbf{0},
\end{aligned}
\end{equation}
which can be used to compute the fidelity and efficiency in the same way as for the PNRD case.

In the absence of detector noise and for a given $\eta$, $F_\text{op}$ can be maximized by increasing $F_\eta$ arbitrarily close to 1 by taking $\vartheta\rightarrow 0$ and sacrificing efficiency. However, detector noise places an additional constraint on the fidelity due to the presence of a finite noise floor. This gives rise to an optimal $\vartheta\neq 0$ that maximizes fidelity. In the regime where $1-\xi_0\simeq\xi_1\ll\eta$, we find that equation (\ref{fgennoise}) for the PNRD case is maximized when $\vartheta\simeq[\xi_1/(\eta(1-\eta))]^{1/4}$. Note that this estimate is also only accurate for $\eta<\xi_0\simeq 1$ as evidently $\vartheta=\pi/4$ is the optimal choice for $\eta=1$ when using a PNRD. As for the BD case, the optimal $\vartheta$ depends on $M_{12}$ due to the contribution from two-photon events. For $M_{12}\simeq 1$ we use the conditional states in equation (\ref{bdnoise}) to find that $\vartheta\simeq [2\xi_1/(\eta(2-\eta))]^{1/4}$ maximizes the fidelity. When $\eta=1$ this optimal choice becomes $\vartheta\simeq(2\xi_1)^{1/4}$. In the regime of quantum communication where $\eta\ll1$, two-photon detections are suppressed due to losses and so the PNRD and BD models give equivalent results.

\subsection{Spin-time bin entanglement (protocol $\mathsf{T}$)}
\label{subsec:time}
For the second protocol (denoted by $\mathsf{T}$), which uses spin-time bin entanglement, we focus on the extension of protocol $\mathsf{N}$ where two successive photons herald entanglement between L-type systems (see figure \ref{fig:Lsystem}). This protocol is also referred to as the Barrett-Kok scheme \cite{barrett2005efficient}, which was utilized to demonstrate the first loophole-free Bell inequality violation \cite{hensen2015experimental}.

\emph{Protocol description.---}Each system is first prepared in the maximal superposition state $(\ket{\uparrow}+\ket{\downarrow})/\sqrt{2}$. Then a resonant  $\pi$ pulse excites the $\ket{\uparrow}$ states at $t_0$, giving equation (\ref{eq:singlephoton_initialstate}). Following protocol $\mathsf{N}$, we could obtain the entangled state $\ket{\psi^{\pm}}$ by post-selecting on a single photon. However, to eliminate the infidelity caused by both systems emitting photons after the first pulse, we can flip the spin state of both systems and re-excite $\ket{\uparrow}$ some time $t_x-t_0$ after the first pulse. If the quantum systems emit only one photon either before or after the second pulse, then they are each in a spin-time bin entangled state $\ket{\psi}_k=(\ket{\downarrow}\ket{\text{early}}+\ket{\uparrow}\ket{\text{late}})/\sqrt{2}$, where $\ket{\text{early}}$ and $\ket{\text{late}}$ represent the presence of a photon in the early  and late time bin modes, respectively. The joint state $\ket{\psi}_1\otimes\ket{\psi}_2$ can be written in the Bell basis of the spin and photon states
\begin{equation}
\label{bellbasis}
\frac{1}{2}\left(\ket{\psi^+}\ket{\psi^+} -\ket{\psi^-}\ket{\psi^-} +\ket{\phi^+}\ket{\phi^+} -\ket{\phi^-}\ket{\phi^-} \right)_\text{s,p},
\end{equation} 
where $\ket{\psi^{\pm}}_\text{s}$ is as before and $\ket{\phi^\pm}_\text{s}=\frac{1}{\sqrt{2}}(\ket{\uparrow\uparrow}\pm\ket{\downarrow\downarrow})$. The Bell states $\ket{\psi^{\pm}}_\text{p}$ and $\ket{\phi^{\pm}}_\text{p}$ are similarly defined using time bin states $\ket{\text{early}}$ and $\ket{\text{late}}$. Interfering the joint state at a beam splitter performs a partial Bell-state measurement (BSM), allowing the identification of $\ket{\psi^+}_\text{p}$ and $\ket{\psi^-}_\text{p}$ from $\ket{\phi^\pm}_\text{p}$. This projects the spin state onto either $\ket{\psi^+}_\text{s}$ or $\ket{\psi^-}_\text{s}$ with a 50\% total probability.

In the absence of spin-flipping decoherence, and with perfect spin-flipping operations, neither quantum system can emit a photon in both the early and late time bins. Thus, when neglecting detector dark counts, the entanglement fidelity is independent of photon losses and the protocol does not suffer from an inherent efficiency-fidelity trade-off. The ramification is that a photon from each emitter must be transmitted to the beam splitter, which reduces the overall protocol efficiency.

\emph{Conditional states.---}
Let $\mathcal{X}(t_\text{x}^\prime,t_\text{x})$ be the superoperator propagator that performs a spin-flip and re-excitation of both systems beginning at time $t_\text{x}$ and concluding at time $t_\text{x}^\prime$. The conditional state $\hat{\rho}_{\mathbf{n}_l,\mathbf{n}_\text{e}}$ given photon counts $\mathbf{n}_\text{e}$ in the early detection window and $\mathbf{n}_l$ in the late detection window is $\hat{\rho}_{\mathbf{n}_l,\mathbf{n}_\text{e}} =\mathcal{W}_{\mathbf{n}_l}\mathcal{X}\mathcal{W}_{\mathbf{n}_\text{e}}\hat{\rho}$, where $\hat{\rho}$ is the state after the first excitation and $\mathcal{W}_{\mathbf{n}_l}$ ($\mathcal{W}_{\mathbf{n}_\text{e}}$) is the conditional propagator for the late (early) time bin detection window dependent on the time ordering $\mathcal{W}_{\mathbf{n}_l}(t_\text{f},r(t_l^\prime),r(t_l),t_\text{x}^\prime)$ ($\mathcal{W}_{\mathbf{n}_\text{e}}(t_\text{x},r(t_\text{e}^\prime),r(t_\text{e}),t_0)$). The detector window duration for the early and late bins are $T_\text{e} = t_\text{e}^\prime-t_\text{e}$ and $T_l=t_l^\prime-t_l$, respectively. 

There are four measurement outcomes that may indicate successful entanglement: $\{\mathbf{n}_l,\mathbf{n}_\text{e}\}=\{(1,0),(1,0)\}$, $\{(1,0),(0,1)\}$, $\{(0,1),(1,0)\}$, and $\{(0,1),(0,1)\}$. For notation convenience, we concatenate the sets of vectors. For example, $\hat{\rho}_{(1,0),(0,1)}=\hat{\rho}_{1001}$. The conditional states are then given by the appropriate conditional propagators $\mathcal{W}_\mathbf{n}$. Using the case 1001 as an example, we have $\hat{\rho}_{1001} = \mathcal{W}_{10}\mathcal{X}\mathcal{W}_{01}\hat{\rho}$, where $\mathcal{W}_{10}$ and $\mathcal{W}_{01}$ are the same as in protocol $\mathsf{N}$. The remaining conditional states can be similarly expressed in terms of $\mathcal{U}_\mathbf{0}$, $\mathcal{S}$, and $\mathcal{X}$ superoperators, although for brevity we do not display them.

\begin{figure}
    \centering
    \includegraphics[scale=0.5]{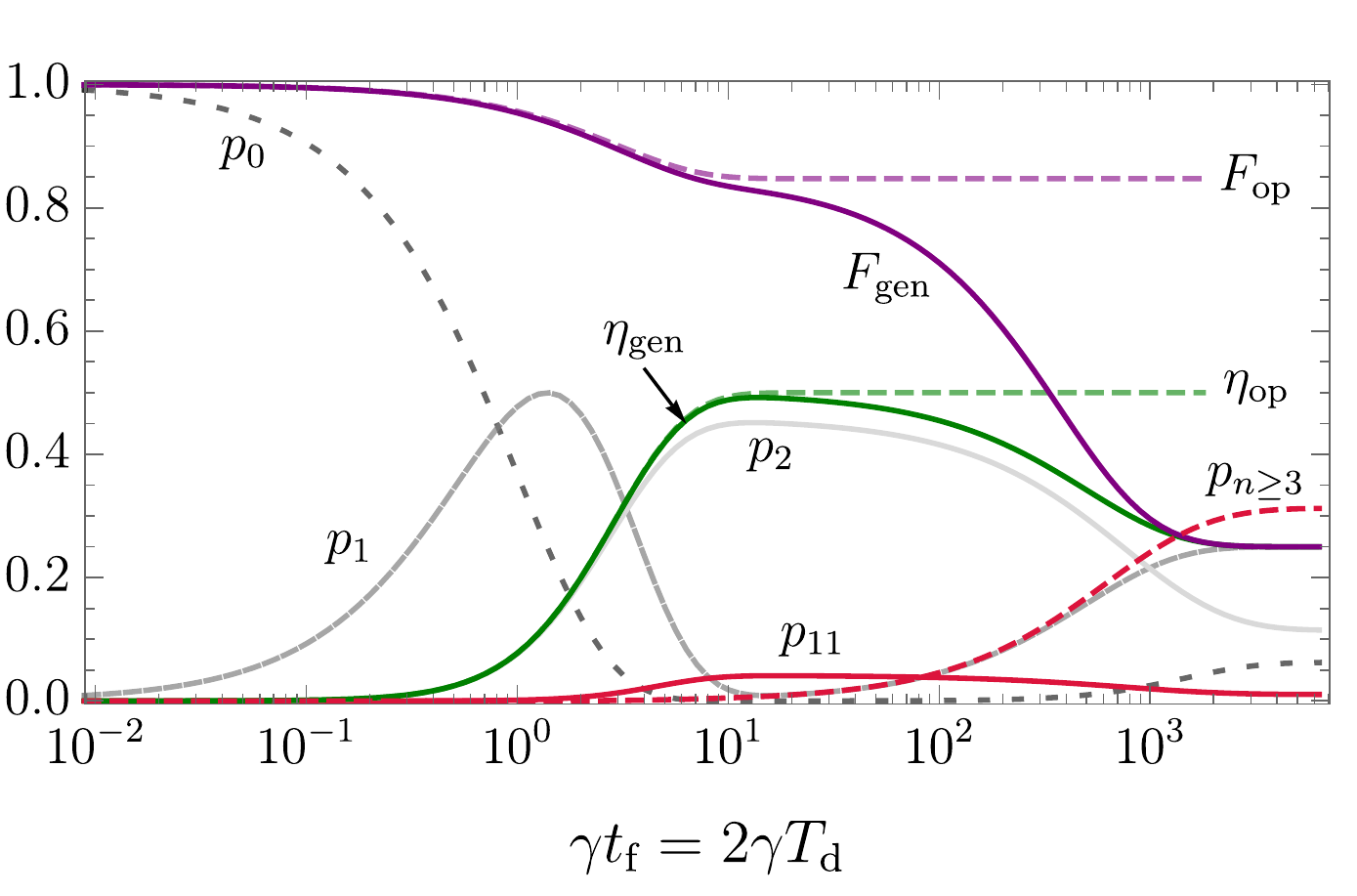}
    \caption{\textbf{Protocol $\mathsf{T}$---time dynamics.} Entanglement generation fidelity $F_\text{gen}$ and efficiency $\eta_\text{gen}$ as a function of protocol time $t_\text{f}$ for an initial state $\hat{\rho}(t_0)=\ket{\psi(t_0)}\!\bra{\psi(t_0)}$ where $\ket{\psi(t_0)}=(1/2)(\ket{\downarrow}+\ket{e})^{\otimes 2}$ and where there is no loss and noiseless local photon-number resolving detectors. The asymptotic dashed lines indicate the limits of fidelity $F_\text{op}$ and efficiency $\eta_\text{op}$ reached when spin decoherence is neglected. The detection window for each time bin is set to be equal to half the protocol duration, which begins after the first system excitation: $t_\text{d}=t_0=0$, $t_\text{d}^\prime=t_\text{x}=T_\text{d}$, and $t_\text{f}=2T_\text{d}$. The gray lines show the probabilities for no photon emission from either system $p_0$, for single photon detection events $p_1$, and for photon bunching events $p_2$. The red solid line shows coincident counts $p_{11}$ caused by imperfect HOM interference and the red dashed line indicates the probability $p_{n\geq 3}$ for 3 or more photons to be emitted as a consequence of spin relaxation between the pulses. Parameters chosen: $\gamma_k=\gamma$, $\gamma_k^\star=0.1\gamma$, and $\gamma_{\text{s}_k}^\pm=\chi_k^\star=0.001\gamma$ for $k\in\{1,2\}$.}
    \label{fig:BK-timedynamics}
\end{figure}

\emph{Measurement duration.---}For protocol $\mathsf{T}$, there are two detection windows beginning at $t_{\text{e}}$ and $t_{\text{l}}$ with duration $T_\text{e}$ and $T_\text{l}$, respectively. Suppose that the detection window is continuous between the pulses. Then we have $t_\text{e}=t_0$, $t_\text{e}^\prime=t_\text{x}$. Also, if the spin-flip and re-excitation is much faster than other system dynamics so that $t_\text{x}^\prime-t_\text{x}\simeq 0$, then we have $t_\text{e}^\prime=t_\text{l}=t_\text{x}$. To simplify the problem, we also make the time bins equal in duration so that $T_\text{e}=T_\text{l}=T_\text{d}$. 

If the detection windows do not encompass the entire photon lifetime, then a high fidelity can be attained because after heralding by two-photon events, both systems will be in the ground state with a high probability. When this is the case, the detection window post selects photons that were emitted early compared to the total lifetime (see figure \ref{fig:BK-timedynamics}). This demonstrates how fast detector gate times can potentially purify photon indistinguishability and increase the overall spin-spin entanglement fidelity. Consequently, the efficiency in this regime is very low. Note that this type of temporal post selection can also be applied to protocol $\mathsf{N}$ provided that $\vartheta$ is very small.

When the time bin duration is on the order of the emission lifetime, the fidelity briefly plateaus at the optical limit where non-zero coincidence counts $p_{11}$ indicate imperfect interference. In this regime, the efficiency approaches the ideal Bell-analyzer efficiency of 50$\%$. However, if the duration is much longer than the optical lifetime, then spin flips occurring between the excitation pulses increase the probability to have three or more photons emitted during the protocol, which reduces the efficiency and fidelity to their thermal limits of $0.25$.

For brevity we do not show the full conditional state solutions for protocol $\mathsf{T}$. However, due to the symmetry of this protocol and its close relationship with protocol $\mathsf{N}$, the fidelity for $\theta=\pi/4$ when neglecting spin decoherence and detector noise takes the simple form
\begin{equation}
\begin{aligned}
\label{BKfidelity}
    F_\text{gen} = \frac{1}{2}\left(1 + \frac{\eta_1\eta_2}{2\eta_\text{gen}}|\tilde{C}(T_\text{d})|^2\right),
\end{aligned}
\end{equation}
where $\eta_\text{gen} =\overline{\beta}_1\overline{\beta}_2/2$ is the efficiency, $\tilde{C}(T_\text{d})$ is given by equation (\ref{P1condsol}), and $\overline{\beta}_k$ is the same as in protocol $\mathsf{N}$. This expression accounts for optical pure dephasing through $\tilde{C}(T_\text{d})$ and can also be averaged for a fluctuating detuning $\Delta$ to capture spectral diffusion.

\emph{Optical limits.---}In the limit that $T_{\text{d}}\gg1/\gamma_k$ we have $\overline{\beta}_k\rightarrow\eta_k$ and $\tilde{C}$ again reduces to equation (\ref{ctilde}). Then the optical limits of efficiency and fidelity are $\eta_\text{op} = \eta^2/2$ and $F_\text{op}=(1+|\tilde{C}|^2)/2$, respectively, for $\eta_k=\eta$. The corresponding concurrence is simply $C_\text{op}=|\tilde{C}|^2$.

As with protocol $\mathsf{N}$, the fidelity and concurrence can be related to the mean wavepacket overlap $M_{12}$ by noting
\begin{equation}
    C_\text{op} = M_{12}\left(\frac{\gamma_1+\gamma_2}{\Gamma_1+\Gamma_2}\right)\leq M_{12}.
\end{equation}
On the other hand, it can be shown that $F_\text{op}\geq M_{12}$. Hence the optical limit of fidelity for $\mathsf{T}$ is bounded by
\begin{equation}
    M_{12}\leq F_\text{op}\leq\frac{1}{2}\left(1+M_{12}\right).
\end{equation}
We note that the upper bound result presented here has also been derived in the supplementary of reference \cite{bernien2013heralded} using arguments from interference visibility.

\emph{Detector noise and number resolution.---}Because of detector dark counts, it is possible that zero or single-photon conditioned states appear to give successful measurements. After taking detector noise into consideration with PNRDs as described in subsection~\ref{measurementsec} we have, for example,
\begin{equation}
\label{BK-PNRD}
\begin{aligned}
\hspace{-1mm}\hat{\varrho}_{1001}= \xi_0^4\hat{\rho}_{1001}
    + \xi_0^3\xi_1\left(\hat{\rho}_{1000}\!+\!\hat{\rho}_{0001}\right)+\xi_0^2\xi_1^2\hat{\rho}_\mathbf{0}.\\
\end{aligned}
\end{equation}

In the absence of detector noise, only conditional states corresponding to three or more total detected photons will cause infidelity when using BDs with protocol $\mathsf{T}$. This only occurs if the probability for a spin flip in between the pulses is non-negligible and photon loss is not too low. Conditional states where two photons arrive at one detector can combine with a single dark count at another detector to cause infidelity. However, for reasonably high photon losses or a reasonably low spin flip probability, both of these contributions to infidelity are negligible compared to other sources. Hence, equation (\ref{BK-PNRD}) also well-approximates the measured state for BDs in this regime. This illustrates the robustness of protocol $\mathsf{T}$ against losses.

\subsection{Spin-polarization entanglement (protocol $\mathsf{P}$)}
\label{subsec:polarization}

\begin{figure}
\includegraphics[width=8.5cm]{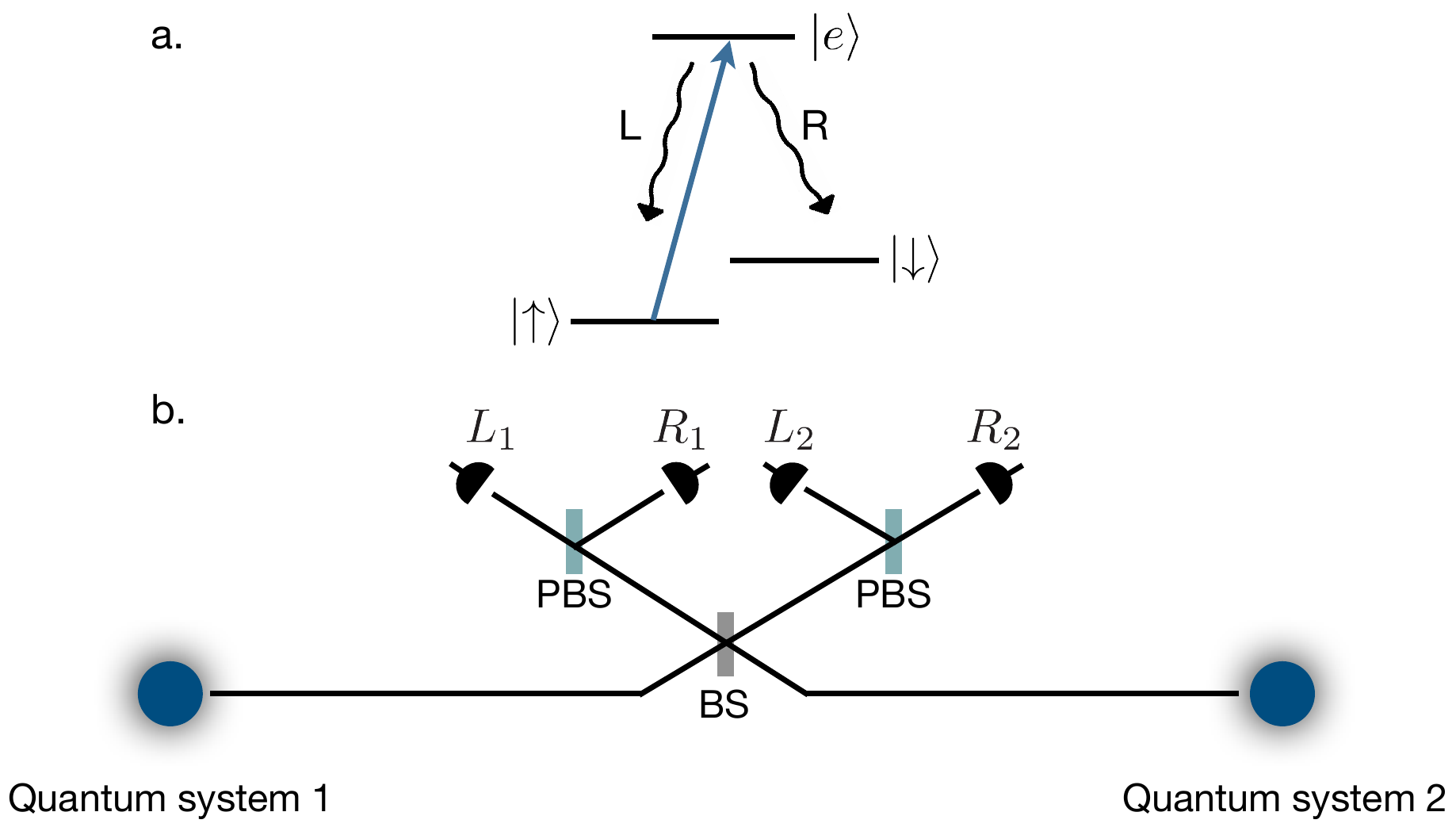}
\caption{\textbf{Entangling pulsed $\mathbf{\Lambda}$ systems.} (a) A $\Lambda$-type system as used in protocol $\mathsf{P}$; an excited state $\ket{e}$ can decay to either ground state $\ket{\uparrow}$ or $\ket{\downarrow}$. (b) Entanglement generation via a polarization Bell-state measurement. Placing $\lambda/2$ and $\lambda/4$ waveplates before each PBS can control the measurement basis.}\label{fig:polarization}
\end{figure}

We now look at the third protocol (denoted by $\mathsf{P}$), which is based on spin-spin entanglement generation via spin-polarization entanglement. For this scheme, we analyze a $\Lambda$-type system where a single excited state $\ket{e}$ can decay to either $\ket{\uparrow}$ or $\ket{\downarrow}$, emitting photons of orthogonal polarization depending on the transition (see figure \ref{fig:polarization}).

\emph{Protocol description.---} Initially, we prepare each of the quantum systems in one of the two ground states. Then, using a short $\pi$-pulse, each system is brought to the excited state. This excited state will then decay to one of the ground states while emitting a photon. 

To illustrate this more clearly, suppose the probability is equal to decay to either ground state. Then the state of the qubit and the emitted photon for each system is $\ket{\psi}_k=\frac{1}{\sqrt{2}}(\ket{\uparrow}\ket{\text{L}}+\ket{\downarrow}\ket{\text{R}})$,
where $\ket{\text{L}}$  and $\ket{\text{R}}$ denote the left and right circular polarization modes of the photon. The joint state of both systems can then be written in the Bell basis for the spin and photon as equation (\ref{bellbasis}), where the polarization modes replace the time bin modes of protocol $\mathsf{T}$.

To perform a BSM, we require a beam splitter (BS) and a polarizing beam splitter (PBS) at each output port of the BS. Then, we place detectors L$_1$ and R$_1$ (L$_2$ and R$_2$) on the left (right) output port of the BS, as shown in figure \ref{fig:polarization}. For $\ket{\psi^+}_{p}$ ($\ket{\psi^-}_{p}$) photon bunching (anti-bunching) happens on the BS. Therefore, considering a perfect interference of the fields, a coincidence in detectors (L$_1$, R$_1$) or (L$_2$, R$_2$) will project the photon state onto the entangled state $\ket{\psi^+}_\text{p}$ and a coincidence in detectors (L$_1$, R$_2$) or (L$_2$, R$_1$) results in the entangled state $\ket{\psi^-}_\text{p}$ \cite{mattle1996dense}. This projects the state of the qubits onto the corresponding spin Bell state. As in protocol $\mathsf{T}$, this setup is not able to distinguish $\ket{\phi^+}_\text{p}$ and $\ket{\phi^-}_\text{p}$ since photon bunching happens for both of these cases. However, with the addition of a source of local auxiliary polarization-entangled photon states, the Bell analyzer success rate could be increased to $75\%$ \cite{grice2011arbitrarily,wein2016efficiency}.

\emph{Conditional states.---}We can describe the source field collected from each transition by $\hat{a}_{\uparrow_k}= \hat{\sigma}_{\uparrow_k}\sqrt{\eta_{c_k}\gamma_{\text{r}\uparrow_k}}$ and $\hat{a}_{\downarrow_k}=\hat{\sigma} _{\downarrow_k}\sqrt{\eta_{c_k}\gamma_{\text{r}\downarrow_k}}$ where $k$ denotes the quantum system 1 and 2 and $\text{r}$ indicates the radiative decay rate. Considering the transmission loss and the beam splitter, we can compute the L-polarized fields $(\hat{d}_1,~\hat{d}_2)$ at detectors L$_1$ and L$_2$ and the R-polarized fields $(\hat{d}_3,~\hat{d}_4)$ at detectors R$_1$ and R$_2$ using equation (\ref{beamsplittereq}). The associated collapse superoperators are then $\mathcal{S}_{i}\hat{\rho} = \eta_\text{d}\hat{d}_i\hat{\rho}\hat{d}_i^\dagger$.

Similar to protocol $\mathsf{T}$, the conditions for a successful protocol are $\mathbf{n} = (1,0,1,0)$, $(1,0,0,1)$, $(0,1,1,0)$, and  $(0,1,0,1)$ where the vectors notate the photon count at the detectors in the order (L$_1$, L$_2$, R$_1$, R$_2$). Like with the previous protocol, we simplify the notation by concatenating the vector elements. In contrast to protocol $\mathsf{T}$, the conditional states for protocol $\mathsf{P}$ are true two-photon events rather than sequential one-photon events. These two-photon conditioned states are computed from their corresponding two-photon conditioned propagators. For example, $\hat{\rho}_{1001}(t_\text{f}) = \mathcal{W}_{1001}(t_\text{f},t_\text{d}^\prime,t_\text{d},t_0)\hat{\rho}(t_0)$ where $\mathcal{W}_{1001}=\mathcal{U}(t_\text{f},t_\text{d}^\prime)\mathcal{U}_{1001}(t_\text{d}^\prime,t_\text{d})\mathcal{U}(t_\text{d},t_0)$ is computed using
\begin{equation}
\begin{aligned}
\label{twophotonpropagator}
    \mathcal{U}_{1001}(t_\text{d}^\prime,t_\text{d})\!&=\!\!\int_{t_\text{d}}^{t_\text{d}^\prime}\!\!\!\int_{t_\text{d}}^{t^{\prime\prime}}\!\!\!\!\mathcal{U}_\mathbf{0}(t_\text{d}^\prime,\!t^{\prime\prime})\mathcal{S}_1\mathcal{U}_\mathbf{0}(t^{\prime\prime}\!,\!t^{\prime})\mathcal{S}_4\mathcal{U}_\mathbf{0}(t^\prime\!\!,t_\text{d})dt^\prime\! dt^{\prime\prime}\\
    &+
    \!\!\int_{t_\text{d}}^{t_\text{d}^\prime}\!\!\!\int_{t_\text{d}}^{t^{\prime\prime}}\!\!\!\!\mathcal{U}_\mathbf{0}(t_\text{d}^\prime,\!t^{\prime\prime})\mathcal{S}_4\mathcal{U}_\mathbf{0}(t^{\prime\prime}\!,\!t^{\prime})\mathcal{S}_1\mathcal{U}_\mathbf{0}(t^\prime\!\!,t_\text{d})dt^\prime\!dt^{\prime\prime}.
\end{aligned}
\end{equation}
Note that by how we defined a photon counting measurement in this work, we are not tracking the arrival time within the detection window. Thus $\mathcal{U}_{1001}$ does not discriminate between the cases where L$_1$ clicks before R$_2$ and cases where R$_2$ clicks before L$_1$. This is illustrated in equation (\ref{twophotonpropagator}) as a consequence of the summation in equation (\ref{conditionalpropagator}). Such a restriction could be lifted if the detectors have sufficient time resolution capabilities.

\begin{figure}
    \centering
    \includegraphics[scale=0.5]{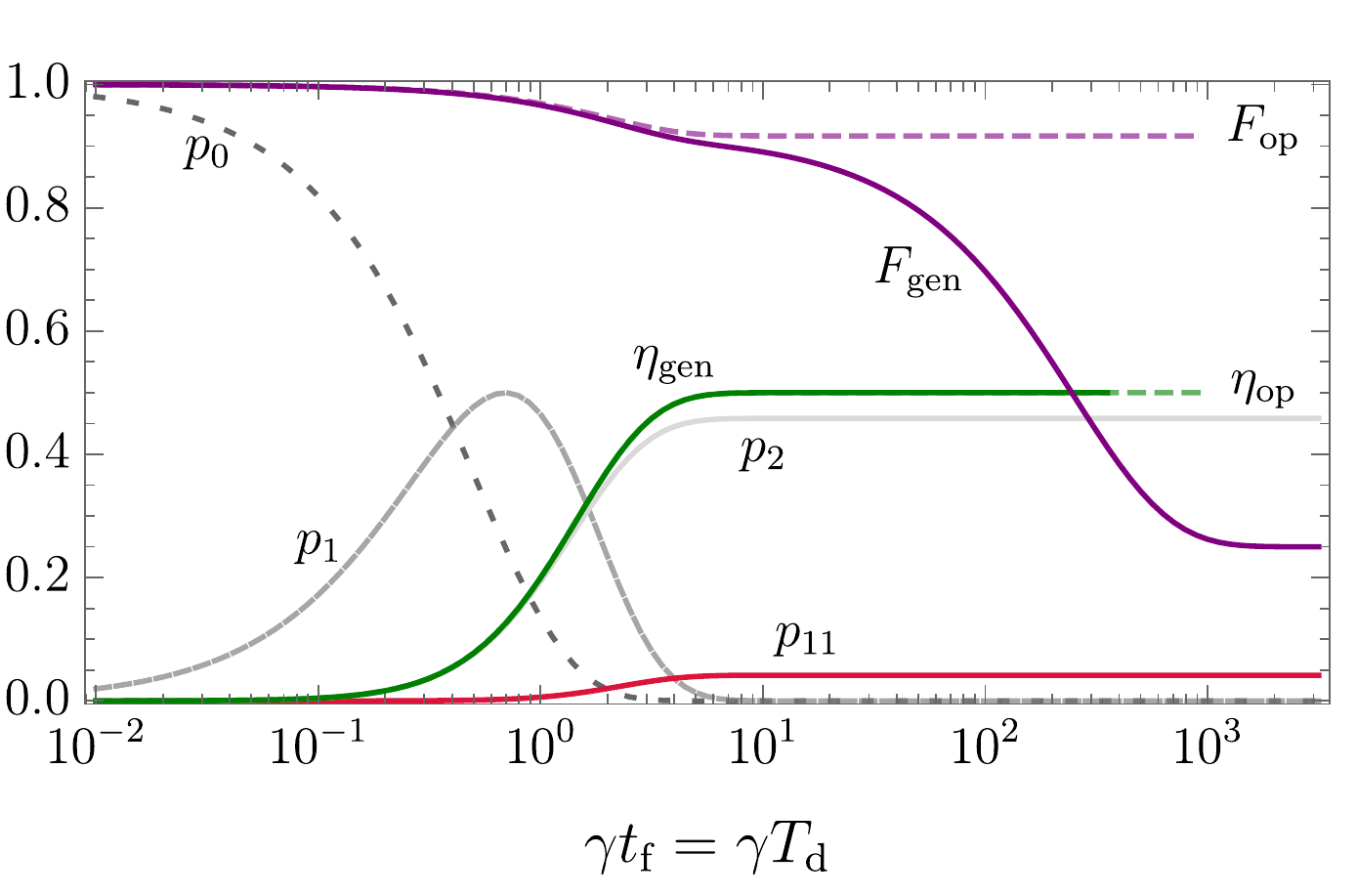}
    \caption{\textbf{Protocol $\mathsf{P}$---time dynamics.} Entanglement generation fidelity $F_\text{gen}$ and efficiency $\eta_\text{gen}$ as a function of protocol time $t_\text{f}$ for an initial state $\hat{\rho}(t_0)=\ket{ee}\!\bra{ee}$ when there is no loss and noiseless local photon-number resolving detectors. The asymptotic dashed lines indicated the limits of fidelity $F_\text{op}$ and efficiency $\eta_\text{op}$ reached when spin decoherence is neglected and when the detection window encompasses the entire photon lifetime. The detection window is set to be equal to the entire protocol duration: $t_\text{d}=t_0=0$ and $t_\text{d}^\prime=t_\text{f}=T_\text{d}$. The gray lines show the probabilities for no photon emission from either system $p_0$, for single-photon events $p_1$ where only one photon is detected, and for photon bunching where two photons arrive at one detector $p_2$. The red line indicates the probability for coincident counts $p_{11}$ caused by imperfect HOM interference. Parameters chosen: $\gamma_{j_k}=\gamma/2$ for $j\in\{\uparrow,\downarrow\}$ and $k\in\{1,2\}$ so that $\gamma_k=\gamma_{\uparrow_k}+\gamma_{\downarrow_k}=\gamma$, $\gamma_k^\star=0.1\gamma$; and $\gamma_{\text{s}_k}^\pm=\chi_k^\star=0.001\gamma$.}
    \label{fig:P3timedynamics}
    \vspace{-1mm}
\end{figure}

\emph{Measurement duration.---} The time dynamics of protocol $\mathsf{P}$ shows features in common with both protocols $\mathsf{N}$ and $\mathsf{T}$. Like $\mathsf{T}$, it is a two-photon scheme and so the fidelity is high for small $T_\text{d}$ compared to the system optical lifetimes $1/\gamma_k$. However, like $\mathsf{N}$, $\mathsf{P}$ only requires a single excitation of each system. Thus the efficiency is unaffected by spin flip processes when $T_\text{d}$ is much larger than the lifetime (see figure \ref{fig:P3timedynamics}).

Although it is possible to derive analytic expressions for protocol $\mathsf{P}$ for arbitrary measurement duration when neglecting spin decoherence, they do not provide new physical insight. For brevity, we only show analytic results in the optical limit to compare with protocols $\mathsf{N}$ and $\mathsf{T}$.

\emph{Optical limits.---}
Consider the case where spin decoherence is negligible, the measurement window encompasses the lifetime $T_\text{d}\gg 1/\gamma_{j_k}$, and the interference is balanced so that $\theta=\pi/4$ and $\eta_{j_k}=\eta$ for $j\in\{\uparrow,\downarrow\}$ and $k\in\{1,2\}$. Then the entanglement generation efficiency is given by $\eta_\text{op}=\eta^2/2$ and the fidelity becomes
\begin{equation}
F_\text{op}=\frac{1}{2}\left(1+\text{Re}\!\left(\frac{\tilde{C}_\uparrow^*\tilde{C}_\downarrow}{\tilde{M}_{\gamma^\star}}\right)\right),
\end{equation}
\vspace{-1mm}
where
\begin{equation}
\begin{aligned}
    \tilde{C}_j &= \frac{2\sqrt{\gamma_1\gamma_2}}{\Gamma_1+\Gamma_2+2i\Delta_j}\\
    \tilde{M}_{\gamma^\star}&=\frac{\gamma_1+\gamma_2-i(\Delta_\uparrow-\Delta_\downarrow)}{\Gamma_1+\Gamma_2-i(\Delta_\uparrow-\Delta_\downarrow)},
\end{aligned}
\end{equation}
and where $\gamma_k=\gamma_{\uparrow_k}+\gamma_{\downarrow_k}$ is the total decay rate of the $k^\text{th}$ system, $\Gamma_k=\gamma_k+2\gamma_k^\star$ is the total optical decoherence rate, $\Delta_\uparrow$ and $\Delta_\downarrow$ are the optical detunings between the left and right circularly polarized transitions (respectively) of the systems. Similar to protocols $\mathsf{N}$ and $\mathsf{T}$, the factor $\tilde{C}_j$ quantifies the coherence for the which-path erasure of photons from the $j$ transitions at the beam splitter, which depends only on the total decay rates relative to the detuning and dephasing. We attribute the factor $1/\tilde{M}_{\gamma^\star}$ to the gain in fidelity due to the systems being initialized in the excited state, compared to protocols $\mathsf{N}$ and $\mathsf{T}$ where the systems are initialized in a superposition state and are directly affected by optical pure dephasing. 

The fidelity is bounded from above by the mean wavepacket overlaps $M_{12\uparrow}$ and $M_{12\downarrow}$ of photons from each transition: \mbox{$F_\text{op}\leq (1+\sqrt{M_{12\uparrow}M_{12\downarrow}})/2$}. Interestingly, this inequality can be saturated if $\Delta_\uparrow-\Delta_\downarrow$ is much smaller than $\gamma_1+\gamma_2$, implying that the systems have nearly identical spin splittings compared to the system decay rate $\gamma_k$. Then we have $\tilde{C}_\uparrow\simeq\tilde{C}_\downarrow$ and $\tilde{M}_{\gamma^\star}\simeq (\gamma_1+\gamma_2)/(\Gamma_1+\Gamma_2)$. In this case, the fidelity becomes $F_\text{op}=(1+M_{12})/2$ where $\Delta = \Delta_\uparrow=\Delta_\downarrow$.

We note that the fidelity of protocol $\mathsf{P}$ does not depend on the ratio of the decay rates to each ground state. Rather, it only depends on the total decay rate $\gamma_k$ of each system, which dictates the photon temporal profile. However, it is still necessary to balance the input intensity at the beam splitter, which may require artificially reducing $\eta_{j_k}$ if some transitions are brighter than others, consequently reducing $\eta$ and the overall efficiency.

\emph{Detector noise and number resolution.---} Because protocol $\mathsf{P}$ is a two-photon heralded scheme, it behaves almost identically to protocol $\mathsf{T}$ in terms of robustness against photon loss and detector noise. This means that the final measured states $\hat{\varrho}_\mathbf{n}$ can be determined using the form of equation (\ref{BK-PNRD}). However, unlike protocol $\mathsf{T}$, protocol $\mathsf{P}$ is quite robust against non-number resolving detectors even when spin flips occur on the order of the emission timescale. This is because spin flips cannot directly affect the photon statistics of protocol $\mathsf{P}$ and so the chance to have more than 1 photon arriving at a given detector remains very small.

\section{Discussion}
\label{sec:discussion}

In this section, we compile the results for each protocol and compare their optical limits of fidelity with respect to each other and the mean wavepacket overlap. We also show the performance of each protocol when including photon loss, detector noise, and spin decoherence.

\subsection{Optical limits}

\begin{figure}
    \centering
    \includegraphics[scale=0.545,trim={6mm 0mm 12mm 10mm},clip]{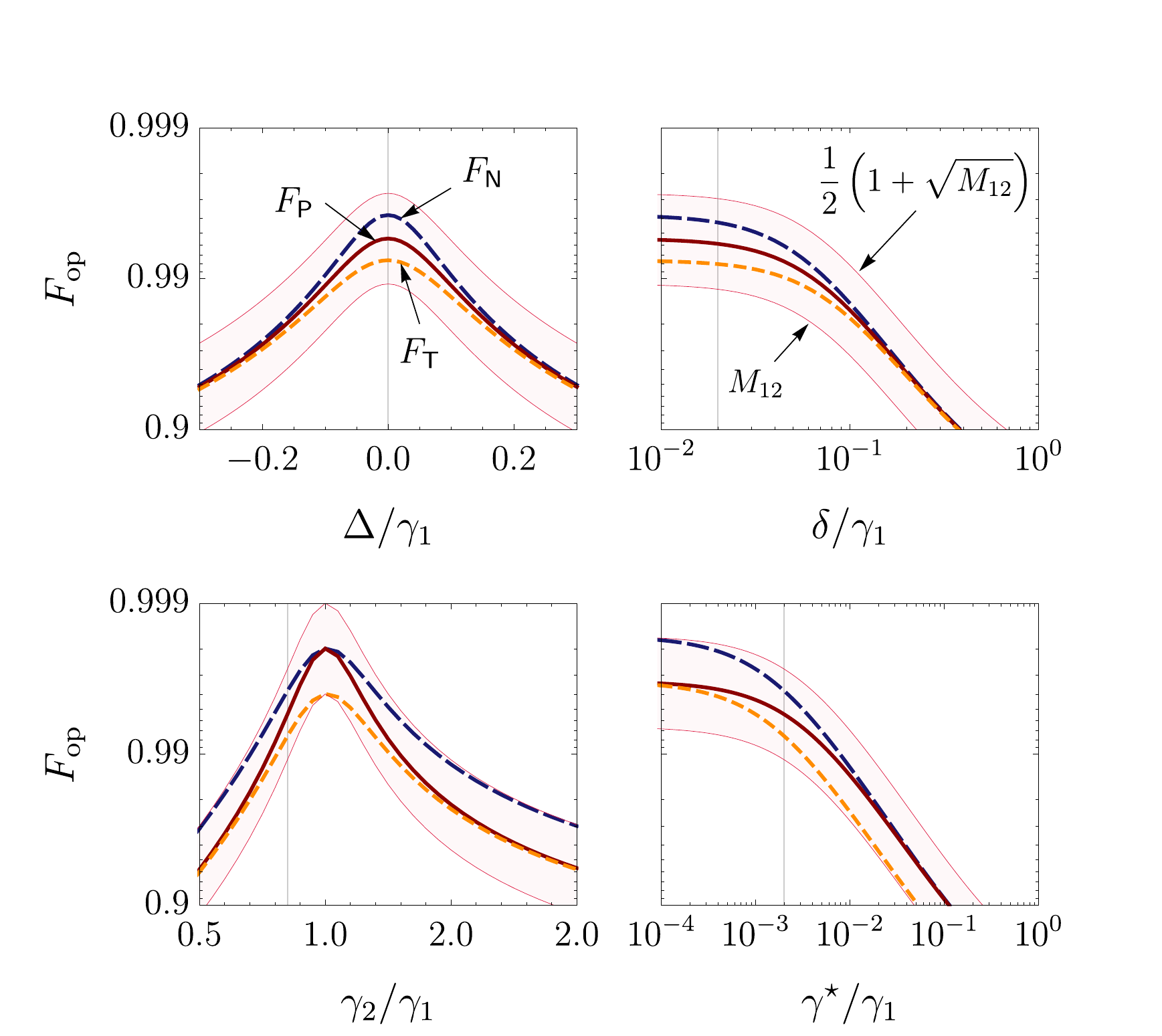}
    \caption{\textbf{Protocol comparison---optical limits.} Fidelity in the optical limit $F_\text{op}$ for each protocol as affected by emitter temporal profile mismatch $\gamma_2/\gamma_1$, pure dephasing $\gamma^\star$, spectral detuning $\Delta$, and spectral diffusion standard deviation $\delta$. These limits are attained when the detection window encompasses the entire photon lifetime and when spin decoherence and loss are negligible. The long dashed blue line represents the fidelity limit for protocol $\mathsf{N}$ when $F_\eta\rightarrow 1$, the short dashed orange line represents protocol $\mathsf{T}$, and the solid red line represents protocol $\mathsf{P}$. The thin red lines and shaded region represent values bounded by the mean wavepacket overlap $M_{12}$ of photons from each source: $M_{12}\leq F_\text{op}\leq (1+\sqrt{M_{12}})/2$. The labeling and order of all lines are the same across all four panels. The thin vertical gray lines in each plot show the fixed values used for each of the other plots. Parameters used unless otherwise stated: $\gamma_2=0.85\gamma_1$, $\gamma^\star_1=\gamma^\star_2=0.002\gamma_1$, $\Delta=0$, and $\delta=0.02\gamma_1$. For protocol $\mathsf{P}$, we also assume that $\Delta_\uparrow=\Delta_\downarrow=\Delta$.}
    \label{fig:comparison-opticallimits}
    \vspace{-4mm}
\end{figure}

In section \ref{subsec:phtnumber}, the maximum fidelity achievable for the spin-photon number entanglement scheme (protocol $\mathsf{N}$) in the limit that $T_\text{d}\gg1/\gamma_k$, $F_\eta(\vartheta)\rightarrow 1$, and $\varphi+\phi=0$ was found to be $F_\text{$\mathsf{N}$} = (1+\text{Re}(\tilde{C}))/2$ corresponding to a concurrence $C_\text{$\mathsf{N}$}=|\tilde{C}|$. The fidelity in section \ref{subsec:time} for the scheme using time-bin entanglement (protocol $\mathsf{T}$) was found to be $F_\text{$\mathsf{T}$} = (1+|\tilde{C}|^2)/2$ corresponding to a concurrence $C_\text{$\mathsf{T}$}=C_\text{$\mathsf{N}$}^2$. In section \ref{subsec:polarization}, we found the fidelity for the spin-polarization entanglement generation scheme (protocol $\mathsf{P}$) to be $F_\text{$\mathsf{P}$} = (1+M_{12})/2$ when $\Delta_\uparrow\simeq\Delta_\downarrow$. The corresponding concurrence is $C_\text{$\mathsf{P}$}=M_{12}$, where $M_{12}$ is the mean wavepacket overlap of photons from each source. 

Knowing that $|\tilde{C}|^2\leq M_{12}$ (see section \ref{subsec:time}) but also $|\tilde{C}|\geq \text{Re}(\tilde{C})\geq M_{12}$ (see section \ref{subsec:phtnumber}), we find that $C_\text{$\mathsf{T}$}\leq C_\text{$\mathsf{P}$}\leq C_\text{$\mathsf{N}$}$. In addition, we have that the order is the same for the fidelity as well: $F_\text{$\mathsf{T}$}\leq F_\text{$\mathsf{P}$}\leq F_\text{$\mathsf{N}$}$. Furthermore, since $F_\text{$\mathsf{N}$}\leq (1+\sqrt{M_{12}})/2$ and $F_\text{$\mathsf{T}$}\geq M_{12}$, the optical limits of fidelity for all three protocols are bounded by
\begin{equation}
    M_{12}\leq F_\text{$\mathsf{T}$}\leq F_\text{$\mathsf{P}$}\leq F_\text{$\mathsf{N}$}\leq \frac{1}{2}\left(1+\sqrt{M_{12}}\right).
\end{equation}

From figure \ref{fig:comparison-opticallimits}, we can see that protocols $\mathsf{N}$ and $\mathsf{T}$ have parallel behaviour in terms of dephasing and temporal overlap due to the fact that $\mathsf{T}$ can be seen as two applications of $\mathsf{N}$. However, protocol $\mathsf{P}$ is implemented with a single pulse on each system like $\mathsf{N}$ but it is still a two-photon scheme like $\mathsf{T}$. Hence it matches the fidelity of $\mathsf{N}$ or $\mathsf{T}$ in different scenarios.

The dominance of protocol $\mathsf{N}$ in the ideal case is expected because the two-photon schemes can naively be seen as two single-photon schemes applied back-to-back, which would compound the infidelity. Because of this, it is tempting to believe that protocol $\mathsf{N}$ would then also be less susceptible to spectral diffusion. However, protocols $\mathsf{T}$ and $\mathsf{P}$ have a symmetry advantage that protocol $\mathsf{N}$ does not have. In protocol $\mathsf{T}$, the fact that the second photon must come from the opposite side of the beam splitter causes an opposing phase rotation on the entangled spin state. These two phases cancel, leaving only a reduction in the magnitude of the coherence due to nonzero $\Delta$ rather than both a reduction and a phase rotation as seen in protocol $\mathsf{N}$. This is illustrated in equation (\ref{BKfidelity}) where the fidelity depends on $|\tilde{C}|^2$ rather than $\text{Re}(\tilde{C})^2$. A similar symmetry occurs for protocol $\mathsf{P}$, however, the detuning phase is only fully eliminated if $\Delta_\uparrow=\Delta_\downarrow$. Because of these symmetry advantages, a sufficient amount of spectral detuning or spectral diffusion eliminates the fidelity advantage that the single-photon scheme had over the two-photon protocols (see figure \ref{fig:comparison-opticallimits}).

\subsection{Phase errors}
\label{phaseerrors}

Let us now discuss the impact of the relative initialization phase $\varphi=\varphi_1-\varphi_2$ and propagation phase $\phi=\phi_1-\phi_2$ errors. The initialization phase $\varphi_k$ for each quantum system can independently fluctuate over time causing significant phase errors if the two quantum systems do not share a phase reference. In addition, it may be necessary to stabilize or correct the propagation phase $\phi$ by monitoring the phase fluctuations of the communication channel \cite{minavr2008phase, yu2020entanglement}. Since the propagation phase depends on distance, this propagation phase fluctuation can become severe for large entanglement generation distances.

As discussed in the previous section, protocols $\mathsf{T}$ and $\mathsf{P}$ have a symmetry advantage over protocol $\mathsf{N}$ for the spectral detuning phase. This advantage extends to propagation phase errors and other possible local phase errors such as initialization phase and the relative precession of the two spin qubits. On the other hand, the upper bound on fidelity for protocol $\mathsf{N}$ can be severely degraded by any phase error $\phi$ becoming $(1/2)(1+\text{Re}(\tilde{C}e^{i\phi}))$. If this phase fluctuates in a Gaussian distribution \cite{minavr2008phase} centered around $\phi=0$ with a standard deviation of $\sigma_\phi$, then the fidelity reduces to $F_\textsf{N}=(1/2)(1+\text{Re}(\tilde{C})e^{-\sigma_\phi^2/2})$ where, as in the previous section, we have assumed $F_\eta\rightarrow 1$.

\begin{figure}
    \centering
    \hspace{-5mm}\includegraphics[scale=0.6]{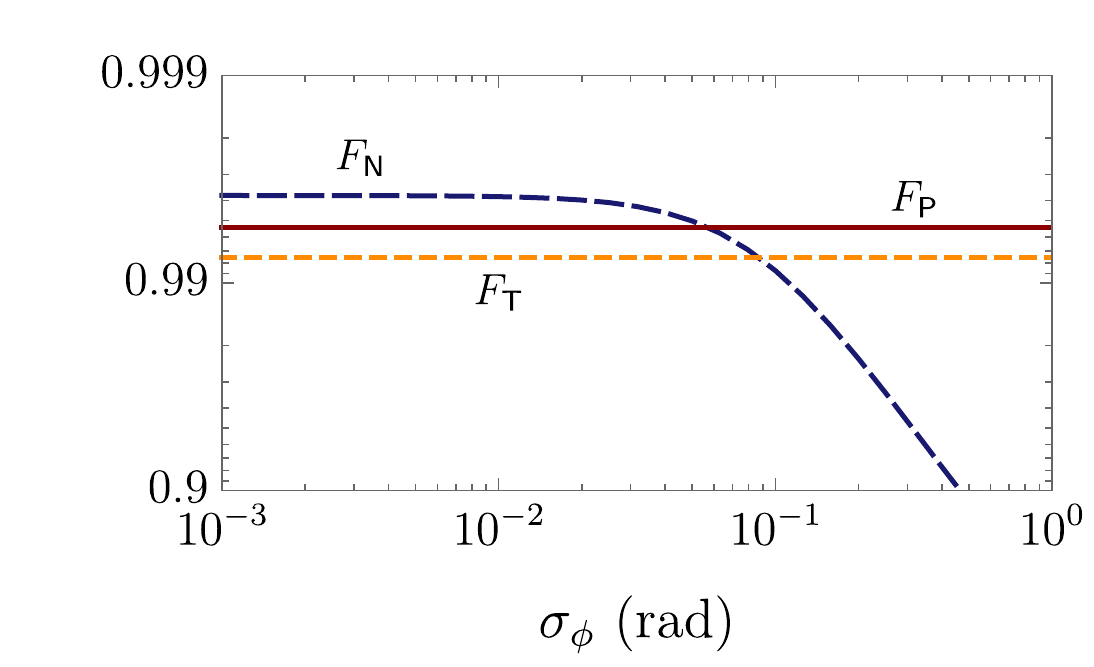}
    \caption{\textbf{Protocol comparison---phase errors.} The reduction in entanglement generation fidelity of protocol $\mathsf{N}$ as compared to phase-robust protocols $\mathsf{T}$ and $\mathsf{P}$ for a phase error $\phi$ fluctuating in a Gaussian distribution around $\phi=0$ with a standard deviation of $\sigma_\phi$. For this comparison, we are neglecting spin decoherence, spectral diffusion, detector dark counts, and detector number resolving limitations. Other parameters used: $\gamma_2=0.85\gamma_1$, $\gamma_1^\star=\gamma_2^\star=0.002\gamma_1$, and $\Delta=0.02\gamma_1$. We also assume that $F_\eta\rightarrow 1$ for protocol $\mathsf{N}$ and that $\Delta_\uparrow=\Delta_\downarrow$ for protocol $\mathsf{P}$.}
    \label{fig:comparisonphaseerror}
\end{figure}

For a large enough phase fluctuation $\sigma_\phi$, protocol $\mathsf{N}$ loses its fidelity advantage over the other two protocols (see figure \ref{fig:comparisonphaseerror}). We find that the value for the variance $\sigma_\phi^2$ where $F_\mathsf{N}\leq F_\mathsf{T}$ is $\sigma_\phi^2\geq \ln((\Gamma_1+\Gamma_2)^2/(4\gamma_1\gamma_2))$. Likewise, for $F_\mathsf{N}\leq F_\mathsf{P}$ we would need $\sigma_\phi^2\geq \ln((\gamma_1+\gamma_2)^2/(4\gamma_1\gamma_2))$.

Although protocols $\mathsf{T}$ and $\mathsf{P}$ are very robust against phase errors, they can still be affected under some conditions. If the phase fluctuation occurs on a timescale faster than the separation between pulses for protocol $\mathsf{T}$, then $F_\mathsf{T}$ can be degraded. This could be accounted for in our method by adding different phases for the second detection window when computing the conditional propagators. In addition, significant birefringence in protocol $\mathsf{P}$, quantified by $\omega_{\text{s}_k}$, can cause a small degradation of $F_\mathsf{P}$ due to propagation phase errors. However, since $\omega_{\text{s}_k}\ll\omega_k$, this effect is orders of magnitude smaller than the degradation experienced by protocol $\mathsf{N}$.

\begin{figure*}[ht]
    \centering
    \hspace{-55mm}(a)\hspace{75mm}(b)\\
    $\hspace{10.mm}
    \overbrace{\hphantom{h\hspace{30.8mm}h}}^\text{
    \normalsize QComm.}\overbrace{\hphantom{h\hspace{22mm}h}}^\text{\normalsize DQC}\hspace{18.1mm}
    \overbrace{\hphantom{h\hspace{26mm}h}}^{\mycom{
    \text{\footnotesize Optically}}{
    \text{\footnotesize Limited}}
    }\overbrace{\hphantom{h\hspace{15.8mm}h}}^{\mycom{
    \text{\footnotesize Spin}}{
    \text{\footnotesize Limited}}
    }\hspace{-0.3mm}\overbrace{\hphantom{h\hspace{7.mm}h}}^{\mycom{
    \text{\footnotesize Noise}}{
    \text{\footnotesize Limited}}
    }$\\\vspace{-4mm}
    \hspace{0.3mm}\includegraphics[width=0.41\linewidth]{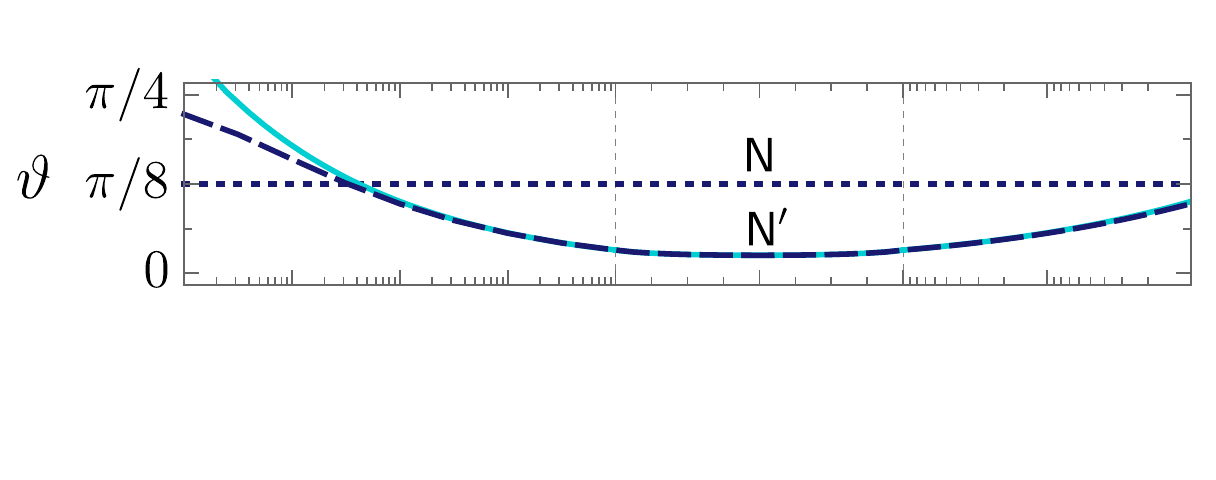}\hspace{5.2mm}
    \includegraphics[width=0.41\linewidth]{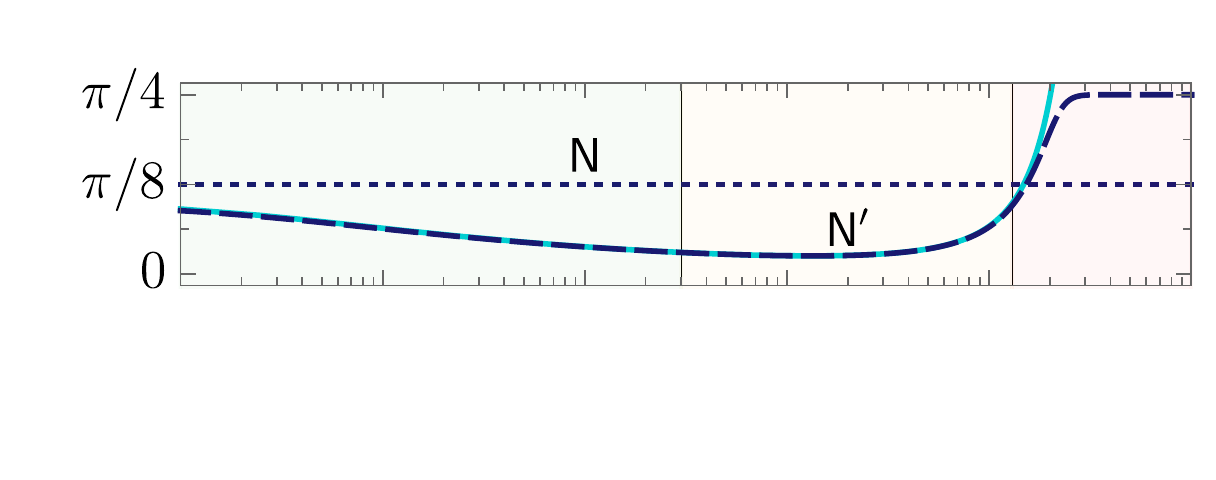}\\\vspace{-13mm}
    \hspace{-3.3mm}\includegraphics[width=0.4281\linewidth]{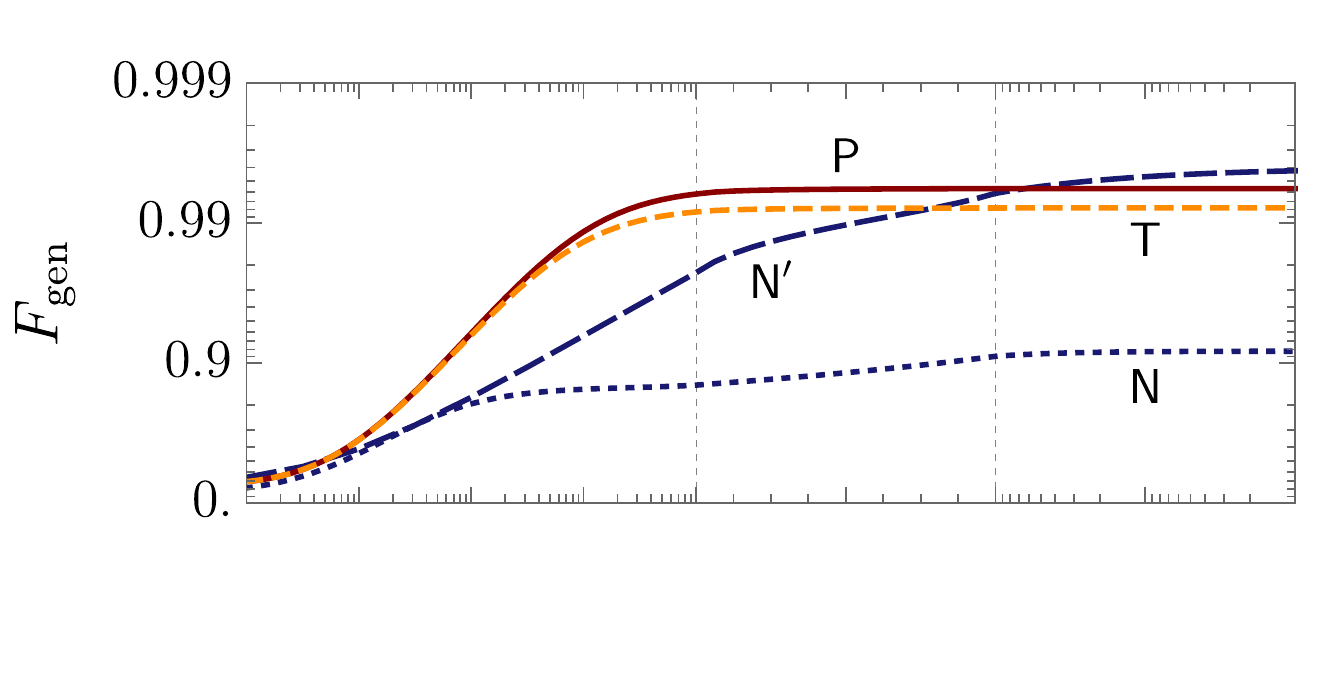}\hspace{1.7mm}
    \includegraphics[width=0.428\linewidth]{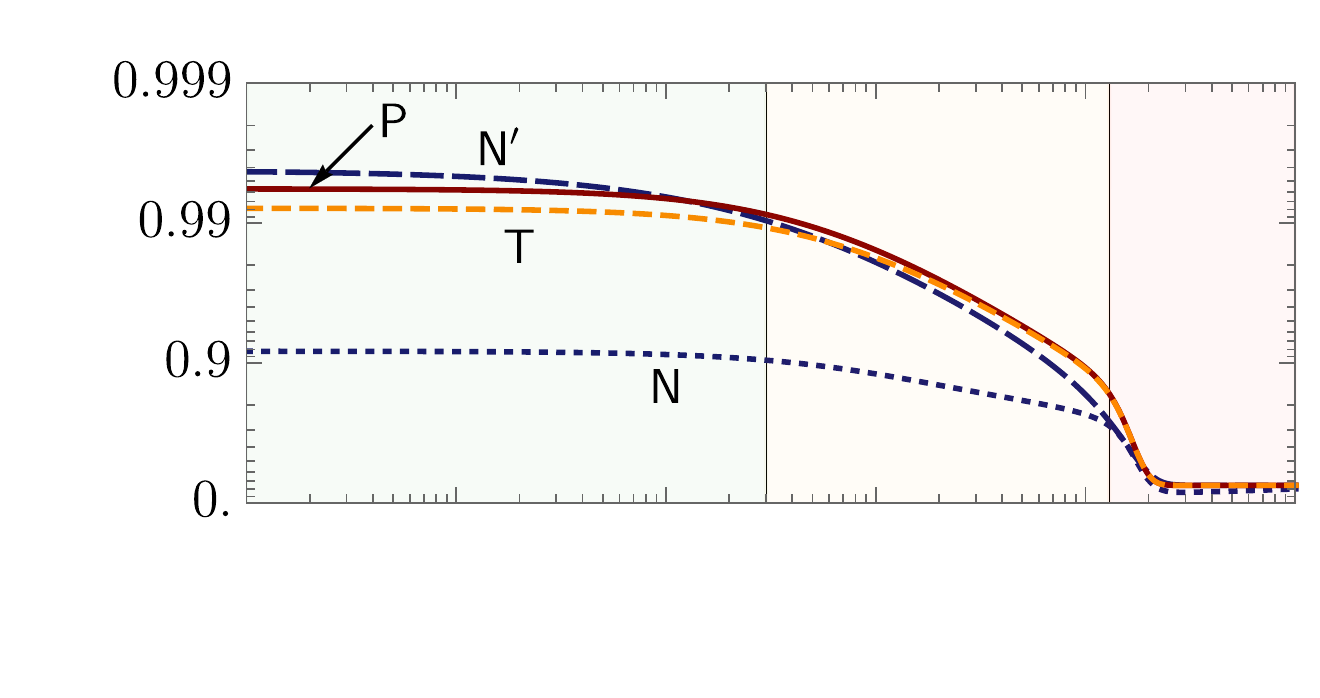}\\\vspace{-13mm}
    \hspace{-1.7mm}\includegraphics[width=0.441\linewidth]{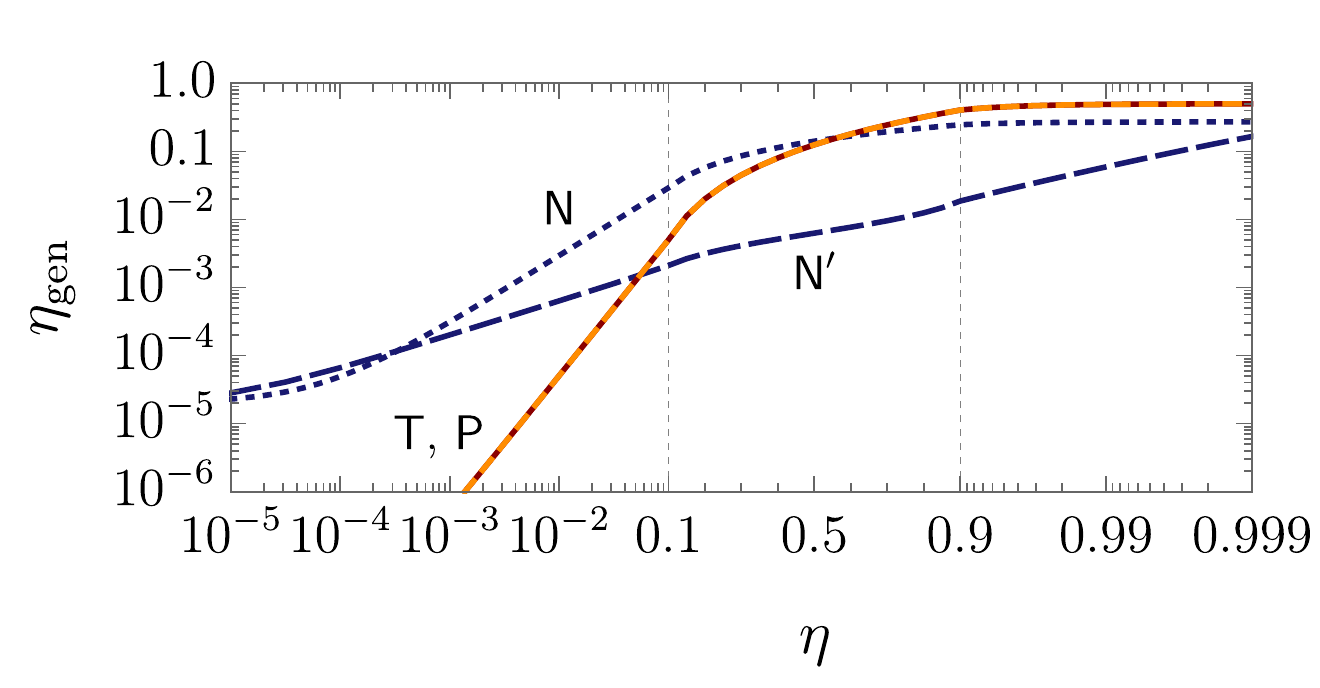}\hspace{-0.1mm}
    \includegraphics[width=0.43\linewidth]{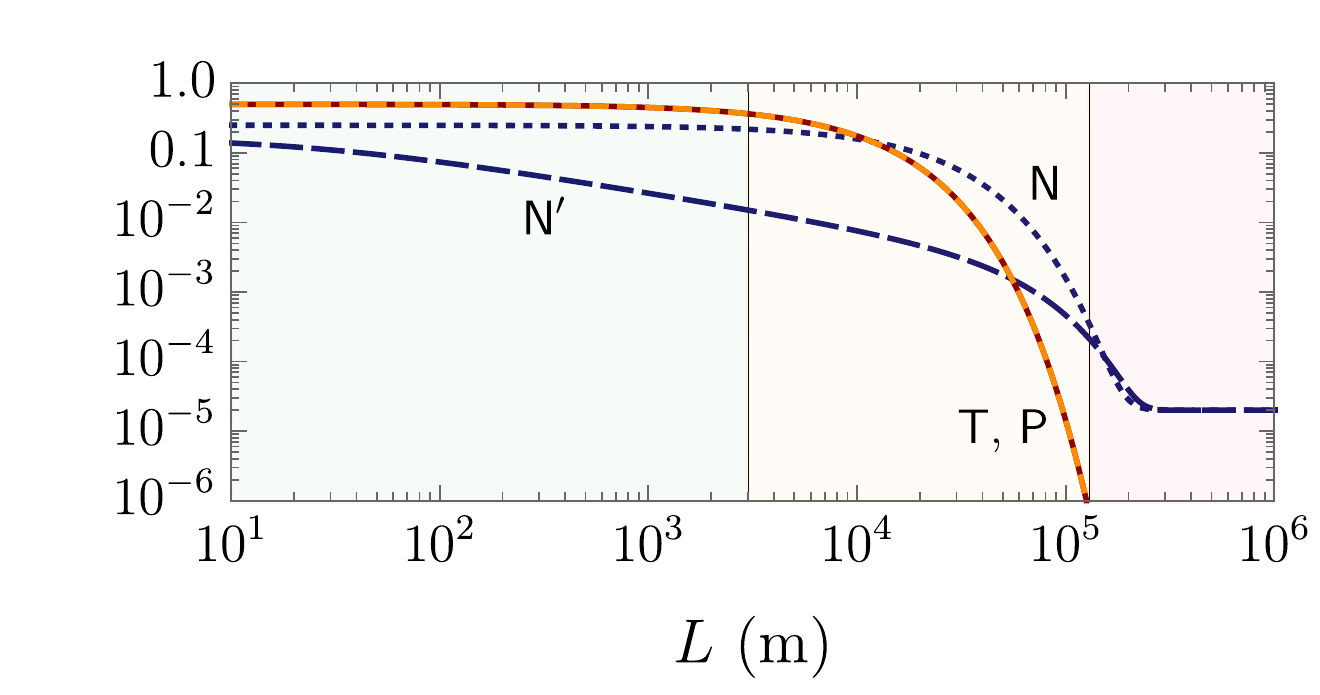}
    \caption{\textbf{Protocol comparison---loss and distance.} (a) A split-scale plot of the fidelity $F_\text{gen}$ and efficiency $\eta_\text{gen}$ as a function of total single-photon efficiency $\eta$ for entanglement generation via spin-photon number entanglement ($\mathsf{N}$), spin-time bin entanglement ($\mathsf{T}$), and spin-polarization entanglement ($\mathsf{P}$) using local detectors with a finite detector dark count probability of $1-\xi_0=10^{-5}$ per detection window of duration $T_\text{d}$. (b) The effect of distance $L$ between the quantum systems on the fidelity and efficiency for each protocol in panel (a) when taking into account spin decoherence in addition to photon loss and detector noise. An increased distance affects both the total protocol duration $t_\text{f}=N_\text{w} T_\text{d}+L/c$ and the single-photon efficiency $\eta=\eta_0 10^{-L/2L_\text{att}}$, where $c=2\times 10^{8}$~m/s is the speed of transmission in the fibre, $N_\text{w}=1$ for $\mathsf{N}$ and $\mathsf{P}$, $N_\text{w}=2$ for $\mathsf{T}$, and $L_\text{att}=22$~km is the fibre attenuation length. (a, b) The dotted blue line represents the protocol $\mathsf{N}$ using non-number resolving detectors (BD) and with a fixed $\vartheta=\pi/8$, corresponding to a probability of 1/4 for a source to emit a photon. The long-dashed blue line indicated by $\mathsf{N}^\prime$ illustrates the noise-limited maximum possible fidelity for $\mathsf{N}$ using photon-number resolving detectors (PNRD). The top panel shows the corresponding numerically optimized $\vartheta$ and the solid light blue line indicates the analytic approximation for the PNRD model: $\vartheta^4=(1-\xi_0)/(\eta(1-\eta))$. The short-dashed orange line and solid red line corresponding to $\mathsf{T}$ and $\mathsf{P}$, respectively, are visually unaffected when accounting for non-number resolving detectors. Other parameters used: $T_\text{d}=5/\gamma_1$, $\gamma_2=0.85\gamma_1$, $\gamma^\star_k = 0.002\gamma_1$, $\Delta=0.02\gamma_1$, $\delta_k=0$, $\gamma_{\text{s}_k}^\pm=0.5\times 10^{-6}\gamma_1$, and $\chi_k^\star=10^{-6}\gamma_1$ for  $k\in\{1,2\}$. For $\mathsf{P}$, we also assume that $\Delta_\uparrow=\Delta_\downarrow=\Delta$. For panel (b) we set $\eta_0=0.999$ for continuity with panel (a) and choose $\gamma_1=10^{8}$~Hz.}
    \label{fig:comparisonLossDistance}
\end{figure*}

\subsection{Loss and distance}
In this section, we compare the fidelity and efficiency of all three protocols while taking into account all imperfections aside from spectral diffusion and phase errors, which were discussed in the previous subsections.

When including losses and detector noise, the two-photon protocols distinguish themselves significantly from the single-photon protocol. Although less flexible, $\mathsf{T}$ and $\mathsf{P}$ are more robust in terms of fidelity than $\mathsf{N}$ (see figure~\ref{fig:comparisonLossDistance}a). However, using PNRDs, $\mathsf{N}$ can exceed $\mathsf{T}$ and $\mathsf{P}$ in terms of fidelity in the regime of distributed quantum computing (DQC) where infidelity may be dominated by optical imperfections. It can also exceed $\mathsf{T}$ and $\mathsf{P}$ in terms of efficiency in the loss regime of quantum communication (QComm). This latter advantage can come at a significant cost to fidelity if there is significant detector noise, even after optimizing $\vartheta$ to minimize the infidelity caused by both systems emitting a photon.

To simulate each protocol's performance over distance, it is necessary take the classical communication time into account as shown in equation (\ref{equ:dmat_dis}), such that the measurement takes place at the retarded time $r(t) = t - L/2c$, where $L=2L_\text{d}$ is the total distance between the systems. The final protocol time also cannot be less than  $t_\text{f} = N_\text{w} T_\text{d} + L/c$, where $N_\text{w}$ is the number of detection windows; $N_\text{w}=1$ for $\mathsf{N}$ and $\mathsf{P}$, $N_\text{w}=2$ for $\mathsf{T}$. This delay caused by the classical communication time can cause a degradation of the entanglement generation fidelity due to spin decoherence.

To compare the protocols, we have selected a set of parameters that best illustrate their differences while also remaining relevant to realistic systems. We have chosen an optical lifetime of 10 ns, with a spin $T_1^\pm$ time of 20 ms and a spin $T_2^\star$ of 10 ms typical of a nitrogen-vacancy center in diamond \cite{fu2009observation,bernien2013heralded}. However, we have chosen an optimistic pure dephasing rate of $0.2$ MHz corresponding to nearly Fourier-transform limited lines, which for many systems would likely require some cavity enhancement or spectral filtering to achieve. In figure~\ref{fig:comparisonLossDistance}b, we set $\eta_0=0.999$ for $L=0$ to illustrate the distance-limited values. In practice, $\eta_0$ is much lower due to other inefficiencies such as collection losses. This may include filtering losses as a consequence of suppressing the excitation laser or phonon sideband emission.

Some differences in fidelity between the protocols seen in figure~\ref{fig:comparisonLossDistance}a are washed out by spin decoherence when the distance approaches or exceeds the fibre attenuation length $L_\text{att}=22$~km. However, the differences in efficiency scaling remain apparent, with $\mathsf{N}$ having the potential to exceed the efficiency of $\mathsf{T}$ and $\mathsf{P}$ by a couple orders of magnitude for long-distance entanglement generation, although with a modest fidelity for our chosen parameter set.

\section{Conclusions}
\label{sec:conclusions}

In this work, we have demonstrated a powerful and intuitive approach based on conditional propagation superoperators to analytically and numerically compute figures of merit for single-photon heralded entanglement generation protocols subject to dephasing. Our method relies on concepts from quantum trajectories and is apt given its resurgence in related techniques for analyzing emitted field states \cite{fischer2018particle,fischer2018scattering,hanschke2018quantum}. Our approach includes a multitude of realistic imperfections that must be considered when developing a platform for quantum information processing based on solid-state emitters. Some of these imperfections may also be relevant for other quantum emitters, such as trapped atoms and ions experiencing excess dephasing processes.

We have provided simple relations to estimate the fidelity and efficiency for three popular entanglement generation protocols. These results are directly useful for developing future proposals for system-specific applications and may also help guide the experimental development of solid-state emitters for quantum information processing. Furthermore, we have used our results to compare these three protocols in order to reveal their strengths and weaknesses in detail.

Although the analysis in this work focused on a simplified three-level model for the quantum systems, our approach can be applied to more complicated systems, such as those in the critical cavity coupling regime \cite{wein2018feasibility}, spin-optomechanical hybrid systems \cite{ghobadi2019progress}, or perhaps emitters in unconventional hybrid cavities \cite{gurlek2018manipulation,franke2019quantization}. It may also prove to be a powerful tool to analyze other photon counting applications when exposed to decoherence process such as novel single-photon interference phenomena \cite{loredo2019generation} or deterministic entanglement generation using feedback \cite{martin2019single}. Moreover, by extending the decomposition and measurements to include detector temporal resolution, the methods presented in this paper may provide a foundation to analyze the effects of decoherence on photon time-tagging heralded measurements.

\vspace{-2mm}
\section*{Acknowledgements}
The authors would like to thank Sumit Goswami, Sourabh Kumar, and H\'{e}l\`{e}ne Ollivier for useful discussions. SCW would also like to thank the GOSS group at the Centre for Nanoscience and Nanotechnology (C2N) in Palaiseau, France, for hosting him during the preparation of this manuscript and for many inspiring discussions on photon statistics. This work was supported by the Natural Sciences and Engineering Research Council of Canada (NSERC) through its Discovery Grant (DG),  Canadian Graduate Scholarships (CGS), CREATE, and Strategic Project Grant (SPG) programs; and by Alberta Innovates Technology Futures (AITF) Graduate Student Scholarship (GSS) program. SCW also acknowledges support from the SPIE Education Scholarship program.
 
\vspace{4mm}
\section*{Author contributions}
SCW and CS conceived the idea. SCW developed the methods. SCW, JWJ, YFW, and FKA performed the analysis and wrote the manuscript. RG and CS provided critical feedback. CS supervised the project and all authors contributed to editing the manuscript.

\vspace{-2mm}
\section*{Disclosures}
The authors declare no conflicts of interests.

\bibliography{bibfile}{}

\end{document}